\documentclass[usenatbib,usegraphicx,usedcolumn]{mn2e}
\usepackage{bm}
\usepackage{amssymb,amsmath}
\usepackage{graphicx}
\usepackage{natbib}
\usepackage{program}
\usepackage{color}
\definecolor{darkgreen}{rgb}{0.15,0.5,0.15}
\definecolor{darkblue}{rgb}{0.15,0.15,0.5}

\usepackage[breaklinks,colorlinks,citecolor=blue]{hyperref}

\title[BECDM haloes]{Galaxy Formation with BECDM: I. Turbulence and relaxation of idealised haloes}

\author[P. Mocz et. al.]{Philip Mocz$^{1}$\thanks{E-mail: pmocz@cfa.harvard.edu (PM)},  
Mark Vogelsberger$^{2}$\thanks{Alfred P. Sloan Fellow}, Victor H. Robles$^{3}$, Jes\'us Zavala$^{4}$, 
\newauthor
Michael Boylan-Kolchin$^{5}$, Anastasia Fialkov$^{1}$, and Lars Hernquist$^{1}$ \\
$^{1}$Harvard-Smithsonian Center for Astrophysics, 60 Garden Street, Cambridge, MA 02138, USA \\
$^{2}$Department of Physics, Kavli Institute for Astrophysics and Space Research, Massachusetts Institute of Technology, Cambridge, MA 02139, USA\\
$^{3}$Department of Physics and Astronomy, University of California, Irvine, CA 92697,USA\\
$^{4}$Center for Astrophysics and Cosmology, Science Institute, University of Iceland, Dunhagi 5, 107 Reykjavik, Iceland \\
$^{5}$Department of Astronomy, The University of Texas at Austin, 2515 Speedway, Stop C1400, Austin, TX 78712-1205, USA
}

\begin{document}

\date{submitted to MNRAS, May 2017}

\pagerange{\pageref{firstpage}--\pageref{lastpage}} \pubyear{2017}

\maketitle

\label{firstpage}

\begin{abstract}
We present a theoretical analysis of some unexplored aspects of relaxed
Bose-Einstein condensate dark matter (BECDM) haloes. This type of ultralight
bosonic scalar field dark matter is a viable alternative to the standard cold
dark matter (CDM) paradigm, as it makes the same large-scale predictions as CDM
and potentially overcomes CDM's small-scale problems via a galaxy-scale de
Broglie wavelength. We simulate BECDM halo formation through mergers, evolved under the Schr\"odinger-Poisson equations. The formed haloes consist of a soliton core supported against
gravitational collapse by the quantum pressure tensor and an asymptotic
$r^{-3}$ NFW-like profile. We find a fundamental relation of the core=to-halo mass with
the dimensionless invariant $\Xi \equiv \lvert E \rvert/M^3/(Gm/\hbar)^2$ or
$M_{\rm c}/M \simeq 2.6  \Xi^{1/3}$, linking the soliton to global halo
properties. For $r \geq 3.5 \,r_{\rm c}$ core radii, we find equipartition
between potential, classical kinetic, and quantum gradient energies. The haloes
also exhibit a conspicuous turbulent behavior driven by the continuous
reconnection of vortex lines due to wave interference. We analyse the
turbulence 1D velocity power spectrum and find a $k^{-1.1}$ power-law. This
suggests the vorticity in BECDM haloes is homogeneous, similar to
thermally-driven counterflow BEC systems from condensed matter physics, in contrast to a
$k^{-5/3}$ Kolmogorov power-law seen in mechanically-driven quantum systems.
The mode where the power spectrum peaks is approximately the soliton width,
implying the soliton-sized granules carry most of the turbulent energy in BECDM
haloes.
\end{abstract}

\section{Introduction}\label{sec:intro}

The $\Lambda$ cold dark matter ($\Lambda$CDM) model has been very successful at
describing the large scale structure of our Universe, including the statistical
properties of the cosmic microwave background (CMB), and the cosmic web of
galaxies across the ages \citep{2005Natur.435..629S}.  State-of-the-art
$\Lambda$CDM simulations -- e.g., \textit{Illustris} \citep{2014Natur.509..177V,
  2014MNRAS.444.1518V}, \textit{Eagle} \citep{2015MNRAS.446..521S},
\textit{Horizon-AGN} \citep{2014MNRAS.444.1453D} -- include complex modelling of
stellar and gas (baryonic) physics that give rise to the galactic population and
provide quantitative predictions over cosmological volumes in the non-linear
density regime of density contrast for essentially the entire range of mass
scales relevant for galaxy formation.

The $\Lambda$CDM model has a nearly scale-invariant spectrum of density
fluctuations with substantial power on small mass scales. While the behavior of
the power spectrum on extremely small scales depends on the specific physics of
the dark matter particles, generic models based on weakly-interacting massive
particles have power spectra that extend without suppression all the way to
Earth masses \citep{green2004}. This feature of CDM models -- significant power
at small scales -- is the source of a number of enduring inconsistencies with
galaxy population statistics observations, including the deficit of dwarf galaxies (the missing satellites
problem; \citealt{klypin1999, moore1999}) and the problem with the abundance of
isolated dwarfs
\citep{2009ApJ...700.1779Z,2011ApJ...739...38P,2015MNRAS.454.1798K}, as well
as the too-big-to-fail problem \citep{boylan-kolchin2011,2012MNRAS.422.1203B}
and the cusp-core problem
\citep{moore1994,flores1994,2004MNRAS.351..903G,2009MNRAS.397.1169D,2010AdAst2010E...5D}.

One widely-considered way to resolve these problems is baryonic feedback, which
can both alter dark matter gravitational potential wells via supernovae or black
hole feedback and prevent the formation of galaxies in low-mass dark matter
haloes \citep{2013ApJ...765...22B,2015MNRAS.454.2092O}. Feedback therefore
provides a natural mechanism for breaking the scale-free nature of the
underlying dark matter density perturbation spectrum.  Given the null detection
of dark matter particles, however, we must also consider the possibility that
the nature of dark matter is different from that of CDM on small scales, meaning
that non-baryonic physics (perhaps in combination with baryonic effects) breaks
the scale-free nature of dark matter.  

The introduction of dark matter self-interactions \citep{2000PhRvL..84.3760S},
for example, can alleviate these small-scale problems
\citep{2012MNRAS.423.3740V,2013MNRAS.431L..20Z,rocha13,2013MNRAS.430.1722V,2014MNRAS.444.3684V,elbert15,2016MNRAS.460.1399V}.  Alternatively, a cutoff in the primordial
power spectrum can be caused by free-streaming in warm dark matter (WDM) models
\citep{bond1982, bode2001}, potentially matching observations better than a pure
CDM model (e.g., see \cite{2016arXiv161109362S} on the missing dwarfs
problem and  \cite{2017MNRAS.468.2836L} on WDM and too-big-to-fail
problem). 
A cutoff in the power spectrum can also be caused by collisional Silk-like
damping if there are significant interactions between dark matter and
relativistic particles in the early Universe
(e.g. \citealt{2002PhRvD..66h3505B,2014PhRvD..90d3524B,2016JCAP...07..013F}). These type of models
offer non-baryonic solutions to the small-scale problems of CDM
(e.g. \citealt{2014MNRAS.445L..31B,2016MNRAS.460.1399V}).

An ultralight bosonic scalar field is a completely different and intriguing
alternative to the CDM paradigm. A bosonic fluid with a particle mass of
$m\sim 10^{-22}~{\rm eV}$ suppresses small-scale structure in the Universe owing
to macroscopic quantum properties (the uncertainty principle) exhibited by the
fluid \citep{2000ApJ...534L.127P,2000PhRvL..85.1158H,lee96}. The typical de
Broglie wavelength for such a particle -- the largest scale at which quantum
mechanical effects will appear -- is $\lambda_{\rm dB}\sim 1~{\rm kpc}$
\citep{2011PhRvD..84d3531C,2014ASSP...38..107S}, similar to the observed sizes
of the stellar distributions in dwarf galaxies
\citep{marsh14,2015MNRAS.450..209B}. This type of fluid has a high temperature
of condensation (critical temperature is $\sim$ TeV due to the high number density of axion particles), which becomes much larger than the mean field
temperature as the Universe expands and cools. The fluid will thus form a
cosmological Bose-Einstein condensate (BEC; \citealt{lundgren10,rob13,matos09},
i.e., a superfluid) in the early Universe. On large scales, however, the scalar field behaves just like
a collisionless self-gravitating fluid, identical to CDM, and is therefore
consistent with modern large-scale cosmological constraints
\citep{matos09,suarez17,bohua14}. We note here that BECDM is often referred to
by a variety of other names, including scalar field dark matter, axion dark
matter, fuzzy dark matter, quantum dark matter, and $\psi$DM (for recent reviews
on BECDM and its astrophysical and cosmological implications, see
\citealt{2014ASSP...38..107S,2017PhRvD..95d3541H,2016PhR...643....1M}).

Aside from
the possible astrophysical relevance of BECDM, theoretical physics offers
motivation for the existence of ultralight scalar-field particles. The axion was
first postulated in Peccei-Quinn theory
\citep{1977PhRvL..38.1440P,1978PhRvL..40..223W}, as an ultralight scalar
particle that resolves the strong CP problem in QCD. String-theory
compactifications also predict a class of ultralight axions
\citep{2010PhRvD..81l3530A}. Such particles are candidates for
BECDM.

Fully cosmological simulations of BECDM are in their infancy, in part owing to
the demanding constraint on spatial resolution required to evolve kpc scale
quantum fluctuations throughout the domain.  \cite{2009ApJ...697..850W}
simulated axion dark matter in 1 Mpc boxes, and \cite{2014NatPh..10..496S}
presented cosmological simulations of BECDM ($2$~Mpc periodic box, dark
matter only) with an adaptive mesh and sufficient resolution to characterise the
universal soliton-cored haloes that form. \cite{2016PhRvD..94d3513S} performed a
detailed investigation of idealised merger simulations, documenting a
comprehensive parameter study of two-soliton interactions in a non-periodic box.
Although there are analytical and numerical solutions with spherical symmetry
that suggest BECDM can solve the CDM's small-scale problems by forming haloes
with central cores (\citealt{sin94,ji94,urena02,guz04,rob12,med15,mat16}, but see \cite{2012MNRAS.427..839S} for discussion of the impact of finite-temperature effects), numerical
simulations with no assumed symmetries are required to confirm these results in
a more realistic scenario. 

One major consequence of the macroscopic quantum mechanical effects in a BECDM
superfluid is that the fluid admits stable, minimum-energy soliton
configurations known to form at the centers of self-gravitating haloes
\citep{gleiser88,seidel94,bal98,guz04,guz06}. These kpc-scale soliton cores
offer one possible solution to the well-known ``cusp-core problem'' of CDM.
Moreover, these solitons are attractor solutions, i.e., they are solutions to
which initially unstable configurations that undergo perturbations will
converge  \citep{sei90,seidel94,lee89,hawley00,gleiser89,chava11,chava16}.
Given the robust stability of the ground state solitons, it is expected that
they survive even after merging with other solitons, which was also noticed in
\citet{2014NatPh..10..496S}.

Given the difficulties associated with numerical simulations of BECDM, most work
on the subject thus far has relied heavily on analytic theory.  Recently,
\cite{2016arXiv160905856G} carried out an analysis to constrain the boson mass $m$
by fitting the luminosity-averaged velocity dispersion of dwarf spheroidal
galaxies (dSphs) using an analytic soliton core dark matter profile. Assuming
radial symmetry and that the halo has not been modified by baryonic physics,
they placed an upper limit on the ultralight boson mass:
$m<0.4\times 10^{-22}~{\rm eV}$ at the $97.5$ per~cent confidence level. 
This is a tighter bound than the earlier study
of dSph constraints \citep{2015MNRAS.451.2479M} based on the ``mass-anisotropy
degeneracy'' in the Jeans equations that leads to biased bounds on the boson
mass in galaxies with unknown dark matter halo profiles. Recently
\cite{2017arXiv170205103U} found that dwarf galaxies in the Milky-Way and
Andromeda are consistent with $m \sim 10^{-21} \rm eV$. The boson mass constraint from
cosmological structure formation requires $m >10^{-23}~{\rm eV}$ to create a
relevant cut-off in the power spectrum on small scales and remain consistent
with large-scale observations \citep{2015MNRAS.450..209B}, such a constraint comes from the Hubble Ultra Deep Field UV-luminosity function and the optical depth to reionisation as measured from
CMB polarisation. 
\cite{2015PhRvD..91j3512H} establishes a constraint of $m >10^{-24}~{\rm eV}$ that comes from CMB temperature anisotropies, which is a more robust lower bound as it depends only on linear physics and less modelling of, e.g., star formation rate and halo mass functions. Lyman-$\alpha$ constraints suggest $m>3.3\times 10^{-22}~{\rm eV}$ at the $2\sigma$ level,
assuming analogies between BECDM and WDM
\citep{2017PhRvD..95d3541H}. This is in moderate tension with the most recent
dSph results of \cite{2016arXiv160905856G} but consistent with
\cite{2017arXiv170205103U}. Numerical simulations of BECDM systems are vital to
make important progress, including testing the assumptions behind the analytic
estimates and possibly ruling out or confirming the BECDM model.  A first
attempt at modelling the Lyman-$\alpha$ flux power spectrum cutoff with
hydrodynamical simulations \citep{2017arXiv170304683I} suggests a constraint of
$m<10^{-22}~{\rm eV}$; however, these simulations do not include the full quantum
effects of BECDM, only its effects on the initial power spectrum.

This paper is the first of a series aimed at quantifying the small-scale effects
of BECDM in a cosmological context. We use idealised numerical
simulations to analyse previously unexplored properties of relaxed/virialized
BECDM haloes that are not covered by analytic work. We also present a numerical
algorithm to simulate BECDM, which will be coupled with baryonic physics in
fully cosmological simulations in future work. We compare the profiles of our
simulated haloes to the analytic soliton relation and also study in more detail
the granular quantum fluctuations that are present in the simulations but absent
in analytical modeling. We show that the source of these fluctuations is quantum
turbulence, an effect that has been seen in non-self-gravitating BEC systems
\citep{2005JPSJ...74.3248K,baggaley2012thermally,2012PhRvL.109t5304B,2016PhR...622....1T}. The
turbulence in our haloes, arising from reconnections of quantum vortex lines
that form during the merging of haloes, is similar to what is found in
dissipationless quantum superfluids with isotropic turbulence.

The manuscript is organised as follows. In Section~\ref{sec:theory} we discuss
the theoretical background and formulation of BECDM haloes. Our code and
simulation setup is described in Section~\ref{sec:sims}.
Section~\ref{sec:profiles} considers the properties of radially-averaged
profiles of virialized structures. Section~\ref{sec:core} presents the scaling
of the resulting soliton core mass and radius with other fundamental parameters
of the system.  Section~\ref{sec:pspec} explores the turbulent properties of
the virialized haloes, with a focus on the velocity power spectrum. Concluding
remarks are given in Section~\ref{sec:conc}.

\section{Theoretical background}\label{sec:theory}

An ultra light scalar field of spin-0 at zero temperature is described in the non-relativistic limit by the Schr\"odinger-Poisson (SP) equations \citep{1990PhRvD..42..384S,sin94,lee96,2000PhRvL..85.1158H,2014ASSP...38..107S}:
\begin{equation}
i\hbar \frac{\partial \psi}{\partial t} = -\frac{\hbar^2}{2m}\nabla^2\psi + m V\psi
\label{eqn:sp}
\end{equation}
\begin{equation}
\nabla^2 V = 4\pi G(\rho-\overline{\rho})
\end{equation}
where the density of the fluid is defined as $\rho=\lvert \psi\rvert^2$ and $\overline{\rho}$ is the mean density. Here $m$ is the mass of the boson, $\psi$ is the wave-function of the particles normalised so its square norm is the density, and $V$ is the gravitational potential. In such a system, all particles share a common wave function $\psi$, hence the physical density of the fluid traces the probability density distribution $\lvert \psi \rvert^2$.

The physical system can be studied in the field or fluid representation \citep{sua11} via the Madelung transformation \citep{1927ZPhy...40..322M,2011PhRvD..84d3531C}:
\begin{equation}
\psi=\sqrt{\rho} {\rm e}^{iS/\hbar}
\,\,\,,\,\,\,
\mathbf{v} = \nabla S/m,
\end{equation}
which yields the fluid representation of the SP equations:
\begin{equation}
\frac{\partial \rho}{\partial t} +
\nabla\cdot (\rho\mathbf{v}) = 0
\end{equation}
\begin{equation}
\frac{\partial\mathbf{v}}{\partial t} 
+ \mathbf{v} \cdot \nabla \mathbf{u}
= - \frac{1}{m}\nabla (Q+V)
\end{equation}
where $Q$ is the quantum potential:
\begin{equation}
Q = - \frac{\hbar^2}{2m} \frac{ \nabla^2\sqrt{\rho} }{\sqrt{\rho}}.
\end{equation}
Equivalently, one can equate
\begin{equation}
-\frac{1}{m}\nabla Q \equiv \frac{1}{\rho}\nabla\cdot\mathsf{p}_{\rm Q}
\end{equation}
in order to define a non-local quantum pressure tensor
\begin{equation}
\mathsf{p}_{\rm Q} = -\left(\frac{\hbar}{2m}\right)^2 \rho \nabla\otimes\nabla \ln\rho
\end{equation}
This quantum pressure tensor offers support against collapse from
self-gravity. The support is non-local, as it depends on the gradient of the
density.

The system has conserved quantities, including the total mass of the system 
\begin{equation}
M = \int\rho \,d^3x,
\end{equation}
and total energy of the system
\begin{eqnarray}
E &=& \int\left[\frac{\hbar^2}{2m^2}\lvert\nabla\psi\rvert^2
+\frac{1}{2}V\lvert\psi\rvert^2 \right]\,d^3x \\
&=&
\int \frac{\hbar^2}{2m^2}(\nabla\sqrt{\rho})^2\,d^3x
+
\int \frac{\rho}{2}v^2\,d^3x
+
\int \frac{\rho}{2}V\,d^3x \\
&=& K_\rho + K_v + W.
\end{eqnarray}
The total (quantum) kinetic energy is $K = K_\rho + K_v$, where $K_v$ is the
classical contribution and $K_\rho$ is the gradient energy due to the quantum
pressure tensor. $W$ is the potential energy. Quantum systems, like their
classical counterparts, also obey a (well-known) virial (Ehrenfest) theorem:
$0 = 2\langle K\rangle + \langle W\rangle$. In addition, the total angular
momentum
\begin{equation}
\mathbf{L} = \int \mathbf{r}\times \rho\mathbf{v}\,d^3x
\end{equation}
is also a conserved quantity. The systems that we consider in our simulations
have no net angular momentum ($\mathbf{L}=\mathbf{0}$).
Note that in most previous analytical works the systems studied are assumed to be spherically symmetric and in equilibrium, with a harmonic time-dependent phase ($S=S(t)$). In general, $S=S(\mathbf{x},t)$, then spatial wave interference may lead to regions of large variations in velocity, which seeds turbulence in the fluid. These nonlinear effects can only be studied with numerical simulations such as the ones presented here.  

The SP equations admit stable soliton solutions, in which the Heisenberg
uncertainty principle/quantum pressure tensor essentially supports the core against
collapse under self-gravity. The soliton core solutions are well-approximated by
\begin{equation}
\rho_{\rm soliton}(r)\simeq
\rho_0
\left[1+0.091\times \left(\frac{r}{r_{\rm c}}\right)^2\right]^{-8}
\label{eqn:fit}
\end{equation}
\citep{2014PhRvL.113z1302S}, where $r_{\rm c}$ is the core radius and $\rho_0$ is the central density given by:
\begin{equation}
\rho_0\simeq 3.1\times 10^{15} 
\left(\frac{2.5\times 10^{-22}~{\rm eV}}{m}\right)^2
\left(\frac{{\rm kpc}}{r_{\rm c}}\right)^4
\frac{M_\odot}{{\rm Mpc}^3}
\end{equation}
The analytic profile fit to the soliton has a flat slope at the center (a
`core'), and approaches a slope of $r^{-16}$ at the outskirts. This can be
compared with the NFW \citep{1996ApJ...462..563N} profile for CDM, which has an $r^{-1}$ cuspy center and a
fall-off going as $r^{-3}$ at large radii.

The SP equations admit a scaling relation, with scaling parameter $\lambda$ \citep{1994PhRvD..50.3655J}:
\begin{equation}
\left\{t,x,V,\psi,\rho\right\} \rightarrow
\left\{\lambda^{-2}\hat{t},\lambda^{-1}\hat{x},\lambda^2\hat{V},\lambda^2\hat{\psi},\lambda^4\hat{\rho}\right\}
\label{eqn:scaling}
\end{equation}
Owing to the scaling, the total mass, energy, and angular momentum of the system transform as 
\begin{equation}
\left\{
M,E,L
\right\} \rightarrow
\left\{
\lambda\hat{M}, \lambda^3\hat{E}, \lambda\hat{L}
\right\}
\end{equation}
Furthermore, the system may also be scaled by boson mass 
$m\to\alpha m$ as:
\begin{equation}
\left\{t,x,V,\psi,\rho\right\} \rightarrow
\left\{\alpha\hat{t},\hat{x},\alpha^{-2}\hat{V},\alpha^{-1}\hat{\psi},\alpha^{-2}\hat{\rho}\right\}
\label{eqn:alphaScaling1}
\end{equation}
\begin{equation}
\left\{
M,E,L
\right\} \rightarrow
\left\{
\alpha^{-2}\hat{M}, \alpha^{-4}\hat{E}, \alpha^{-3}\hat{L}
\right\}.
\label{eqn:alphaScaling}
\end{equation}

Since we assume no net angular momentum in our simulations, the system is then primarily characterised by a single invariant,
$\lvert E \rvert/M^3$ \citep{2016PhRvD..94d3513S}, which is unchanged under the scaling symmetry.
To make it dimensionless, we define 
the invariant quantity as:
\begin{equation}
\Xi \equiv \lvert E \rvert/M^3/(Gm/\hbar)^2.
\end{equation}
Note that our definition for $\Xi$ is invariant not just under the $\lambda$
scaling of the SP equations but also with the boson mass
(Eqn.~\ref{eqn:alphaScaling1}), which makes our results even more general and scalable to any boson mass. 

BEC superfluid systems (such as superfluid liquid helium) are known to exhibit
turbulent behavior in a number of regimes
\citep{2005JPSJ...74.3248K,baggaley2012thermally,2012PhRvL.109t5304B,2016PhR...622....1T}. Turbulence
in BEC systems is a young and developing field (for a recent review see
\citealt{2016PhR...622....1T}). Many such systems are described by the
Gross-Pitaevskii equations, which are the Schr\"odinger equations with a
non-linear self-interaction term, and no self-gravity (instead, a static
potential `trap' is often assumed). Turbulence is possible due to the advective term
$\mathbf{u}\cdot\nabla\mathbf{u}$ in the fluid formulation of the governing
equations, which is also the case in classical fluid dynamics. 

Quantum
turbulence is different from its classical manifestation, however \citep{baggaley2012thermally,2012PhRvL.109t5304B,2016PhR...622....1T}. A direct
consequence of the definition of the fluid velocity $\mathbf{v}=\nabla S/m$ is
that the flow is irrotational: $\nabla\times\mathbf{v}=0$. However, there is an
\textit{exception} if $\psi$ is not continuous or does not have first or second
derivatives. Vorticity in a quantum fluid is thus restricted to degenerate
vortex lines or cores, where the velocity may diverge to infinity but the
density is zero thus the solution remains physical. These filamentary vortex
line structures naturally reconnect and create Kelvin waves that mediate the
cascade of energy to smaller scales \citep{2016PhR...622....1T}. Vortex
generation is possible from configurations initially smooth and without vortices
\citep{galati2013nonlinear}, but a Kelvin's conservation of circulation theorem
applies to the system, placing topological constraints on the vortex lines that
are allowed to form. 

Turbulence in Gross-Pitaevskii fluids is known to develop Kolmogorov-like
$k^{-5/3}$ velocity power spectra when the fluid is mechanically driven on the
largest scale of $k$ \citep{baggaley2012thermally} (mechanically driven here refers to driving the BEC fluid by grids or propellers): in this case turbulence arising from large spatial-scale modes cascades to smaller scales. In contrast, the spectrum of
thermally-driven turbulence in a BEC fluid (i.e., driven by a small heat flux which introduces a counterflow velocity in the superfluid. There is no stirring length-scale introduced into the problem), which lacks energy on the largest
scales, exhibits a `bump' in the velocity power spectrum at intermediate scales (the inter-vortex lengthscale)
and has been shown to scale as $k^{-1}$ at large wave numbers
\citep{baggaley2012thermally,2012PhRvL.109t5304B,2016PhR...622....1T}. \cite{2011JPhB...44k5101C} also find $k^{-1}$ steady-state power-spectrum in the linear Schr\"odinger equations on small scales due to vortex reconnection.  The connection between
classical and quantum turbulence is an interesting and growing field
\citep{2016PhR...622....1T}. Although understanding the various sources of turbulence is a challenging task, 
it has been noticed that when the turbulence arises from small to large scale the velocity spectrum is $k^{-1}$ which is characteristic of fluid with isotropic turbulence. 

As some of the behavior of superfluids can be quite complicated to capture analytically, especially turbulence, we rely on numerical simulations to study BECDM haloes as described in the next section.

\begin{figure*}
\begin{center}
\begin{tabular}{cc}
$t=0$ & 
$t=0.1t_{\rm H}$ \\
\includegraphics[width=3.3in,angle=0,clip=true] {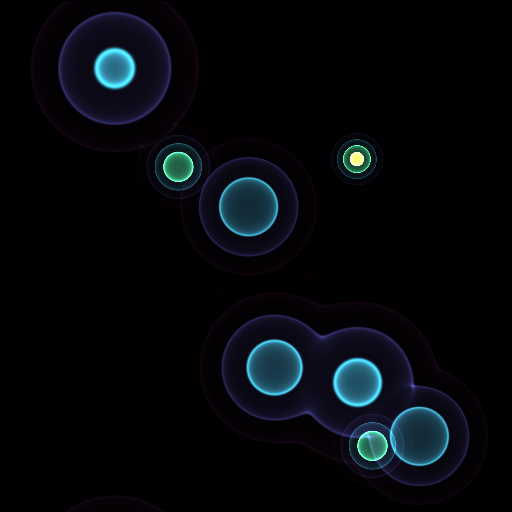}
\llap{\includegraphics[width=3cm]{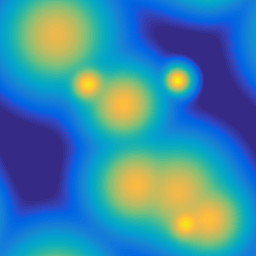}\hspace{5.47cm}} 
& 
\includegraphics[width=3.3in,angle=0,clip=true] {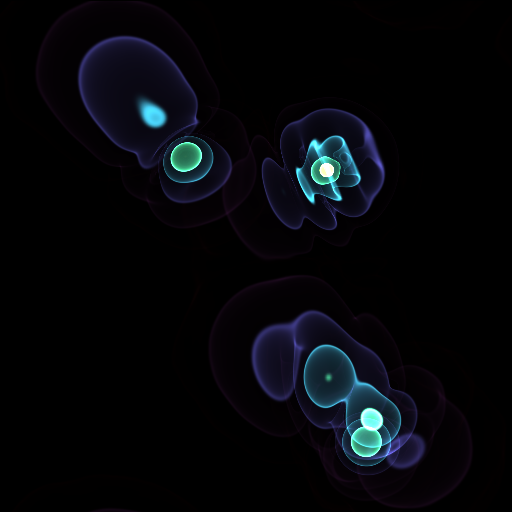}
\llap{\includegraphics[width=3cm]{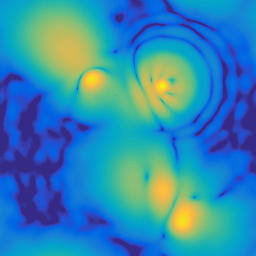}\hspace{5.47cm}}  
\\
$t=0.2t_{\rm H}$ & 
$t=t_{\rm H}$ \\
\includegraphics[width=3.3in,angle=0,clip=true] {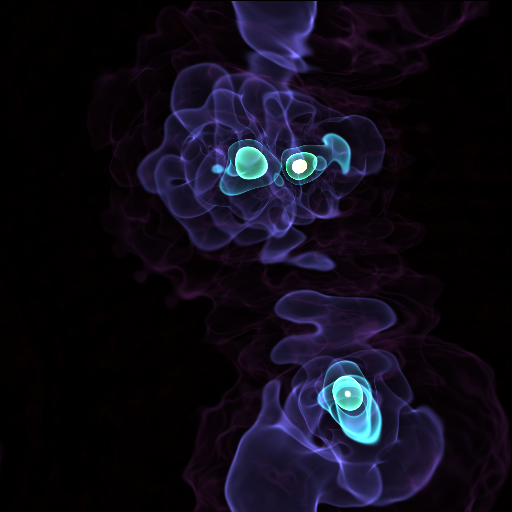}
\llap{\includegraphics[width=3cm]{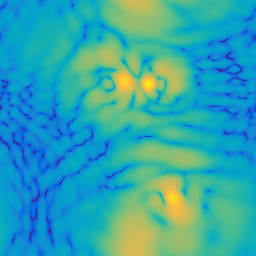}\hspace{5.47cm}} 
& 
\includegraphics[width=3.3in,angle=0,clip=true] {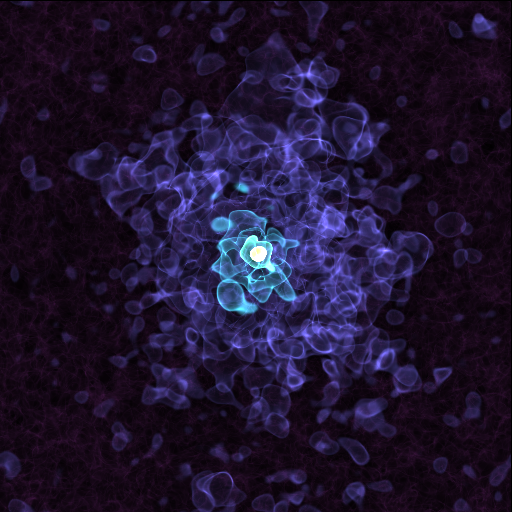} 
\llap{\includegraphics[width=3cm]{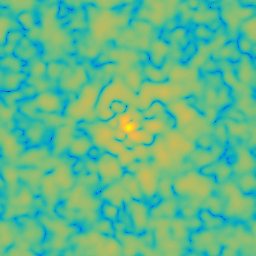}\hspace{5.47cm}}
\end{tabular}
\includegraphics[width=6.6in,angle=0,clip=true] {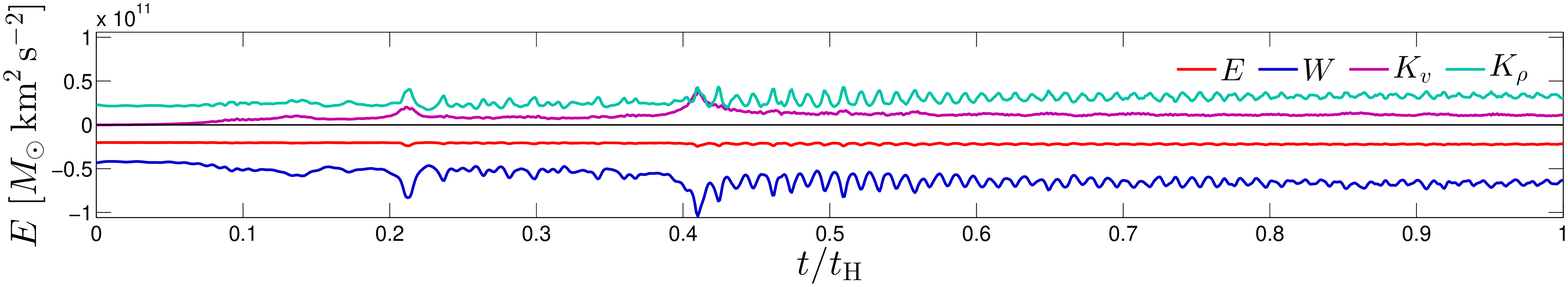}
\end{center}
\caption{Volume rendering of the density field in one of our simulations of the formation of a virialized BECDM halo through multiple mergers. We merge isolated soliton cores ($t=0$) until a single bound halo forms, which is characterised by a stable soliton core at the center of the halo and quantum fluctuations throughout the domain. The volume rendering shows isocontours of density differing by factors of $10$. Insets show projected density in log-space. The \textit{bottom panel} shows the time evolution of the total energy $E$, potential energy $W$, classical kinetic energy $K_v$, and quantum gradient energy $K_\rho$ in the simulation.} 
\label{fig:snaps}
\end{figure*}

\section{Simulations}\label{sec:sims}

We simulate $100$ scenarios of a group of soliton cores that merge to form a final virialized halo to study the general properties of virialized BECDM structures that are statistically meaningful. The simulation setup is based on idealised simulations found in \cite{2014PhRvL.113z1302S,2016PhRvD..94d3513S}, which suggest that the initial conditions to form virialized haloes are largely unimportant for the final outcome. 
In our simulations we mainly create a virialized halo by merging soliton cores of different initial sizes and central densities, but we have also verified, using additional simulations, that we produce consistent types of cored-rather-than-cuspy final BECDM virialized haloes if we merge initially cuspy NFW profiles. Therefore, the simulations are a very useful tool to study the final product of the relaxation of BECDM haloes. The simulations are evolved with a pseudo-spectral method for solving the SP equations in 3-D. Details of the numerical method are provided in the following subsection.

\subsection{Pseudo-Spectral Method}\label{sec:code}

We developed a second-order pseudo-spectral solver for the SP equations, which
we have also added (Mocz et al. in prep.) into the {\sc arepo} code
\citep{2010MNRAS.401..791S}. Our approach employs a `kick-drift-kick'
techniques, akin to symplectic leapfrog $N$-body solvers
\citep{2005MNRAS.364.1105S}. The wavefunction $\psi$ is evolved with unitary
`kick' and `drift' operators.

The variables $\psi(x)$, $\rho(x)$, $V(x)$, are discretized onto a grid of
dimension $N^3$. In this subsection, the variables will represent the
discretized grid versions. Given the density field $\rho$, the potential $V$
can be calculated by transforming to Fourier-space and back:
\begin{equation}
V = 
{\rm ifft}\left[
-{\rm fft}\left[
4\pi G (\rho-\overline{\rho})
\right]/ k^2
\right]
\end{equation}
where ${\rm fft}\left[\cdot\right]$
and ${\rm ifft}\left[\cdot\right]$ are
Fourier transform and inverse Fourier transform operators, respectively,
and $k$ are the wave numbers at the corresponding grid locations.

First, the wavefunction is given a `kick' by half a timestep, due to the potential:
\begin{equation}
\psi \leftarrow \exp\left[-i (\Delta t/2)(m/\hbar)V\right]\psi
\label{eqn:kick}
\end{equation}

This is followed by a full `drift' (kinetic) step in Fourier-space:
\begin{equation}
\hat{\psi} = {\rm fft}\left[ \psi \right]
\end{equation}
\begin{equation}
\hat{\psi} \leftarrow 
\exp\left[ -i \Delta t (\hbar/m) k^2/2  \right]\hat{\psi}
\label{eqn:drift}
\end{equation}
\begin{equation}
\psi \leftarrow  {\rm ifft}\left[ \hat{\psi} \right]
\end{equation}

The timestep is completed with another `kick' step (Eqn.~\ref{eqn:kick}), and
the system is thus evolved from time $t$ to time $t+\Delta t$.

The valid timestep criterion for stability and accuracy of our method,
essentially a Courant-Friedrichs-Lewy (CFL) like condition, is that the unitary
operators in Equations \ref{eqn:kick} and \ref{eqn:drift} do no change the
phase by more than $2\pi$ in each timestep.  The timestep criterion of
\cite{2016PhRvD..94d3513S}:
\begin{equation}
\Delta t \leq {\rm max}\left[
\frac{m}{6\hbar} (\Delta x)^2, 
\frac{h}{ m \lvert V \rvert_{{\rm max}} }
\right]
\end{equation}
enforces this property ($\lvert V \rvert_{{\rm max}}$ is the maximum of the
absolute value of the gravitational potential). Note the timestep scales as
$(\Delta x)^2$ rather than $\Delta x$ for gravity and Eulerian fluid solvers,
which adds computational cost to the simulations.

We briefly compare the differences in existing codes that have been used to
solve the SP equations to simulate BECDM. The advantages of our method include:
simplicity, use of unitary operators, a `kick-drift-kick' formulation which
makes the method readily integrable into a number of existing cosmological
codes, and machine precision control of the total kinetic energy during the
drift step.  This pseudo-spectral method achieves spectral (exponential)
convergence in space and second-order convergence in time. The main limitation
of pseudo-spectral methods in general is their restriction to a regular grid.
Because of this, \cite{2014NatPh..10..496S} solve the SP equations on an
adaptively refined mesh to achieve high dynamic range in cosmological
simulations. The implementation requires Taylor expansion of the unitary
operators with modified coefficients in order to minimise the small-scale
numerical damping.  \cite{2016PhRvD..94d3513S} use a 4th-order Runge-Kutta
finite-difference solver on a grid to solve the SP equations.  In general, our
code yields consistent results with those obtained in alternative codes in the
comparable regime.

\begin{figure}
\begin{center}
\includegraphics[width=0.47\textwidth,angle=0,clip=true] {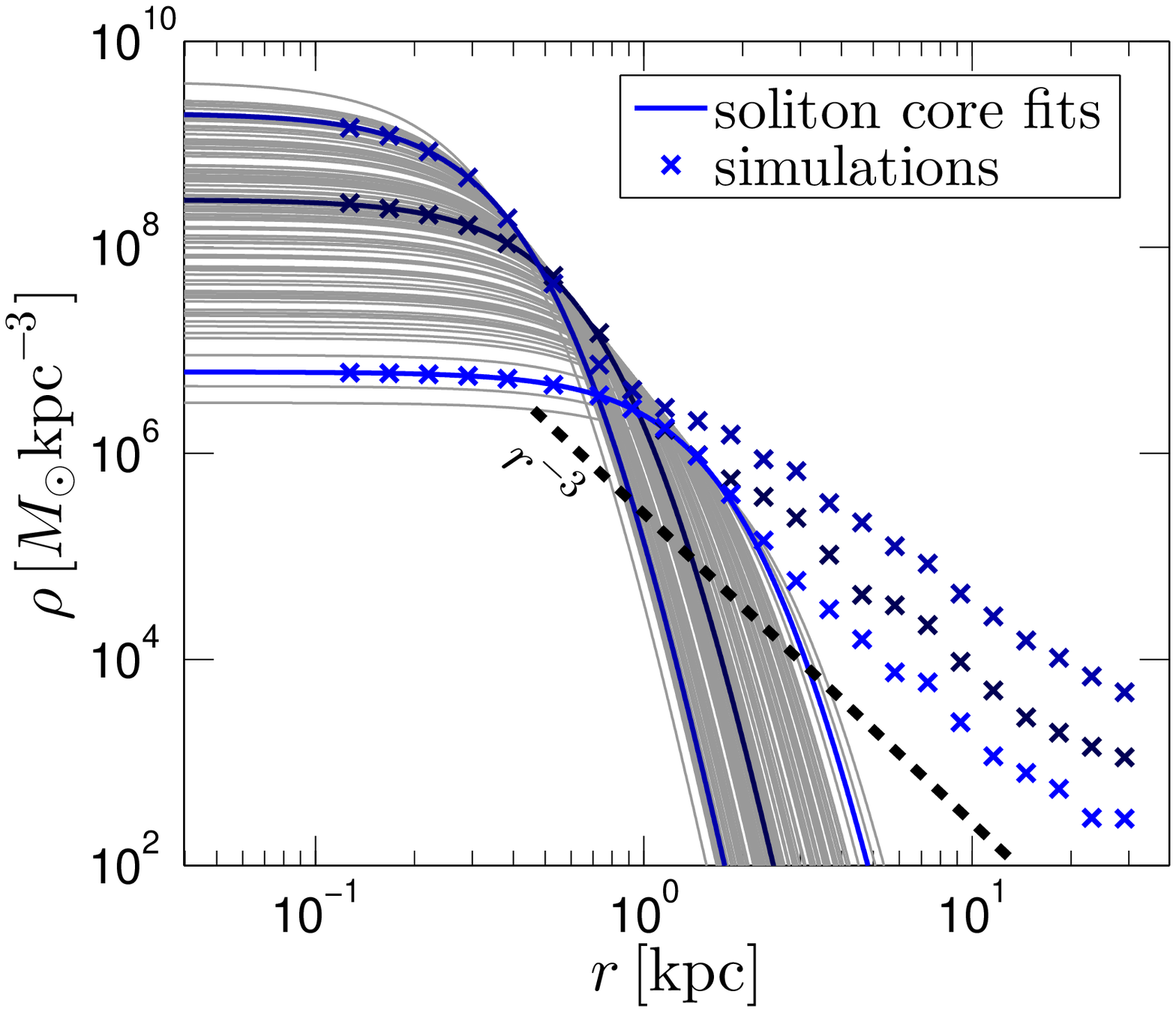} 
\includegraphics[width=0.47\textwidth,angle=0,clip=true] {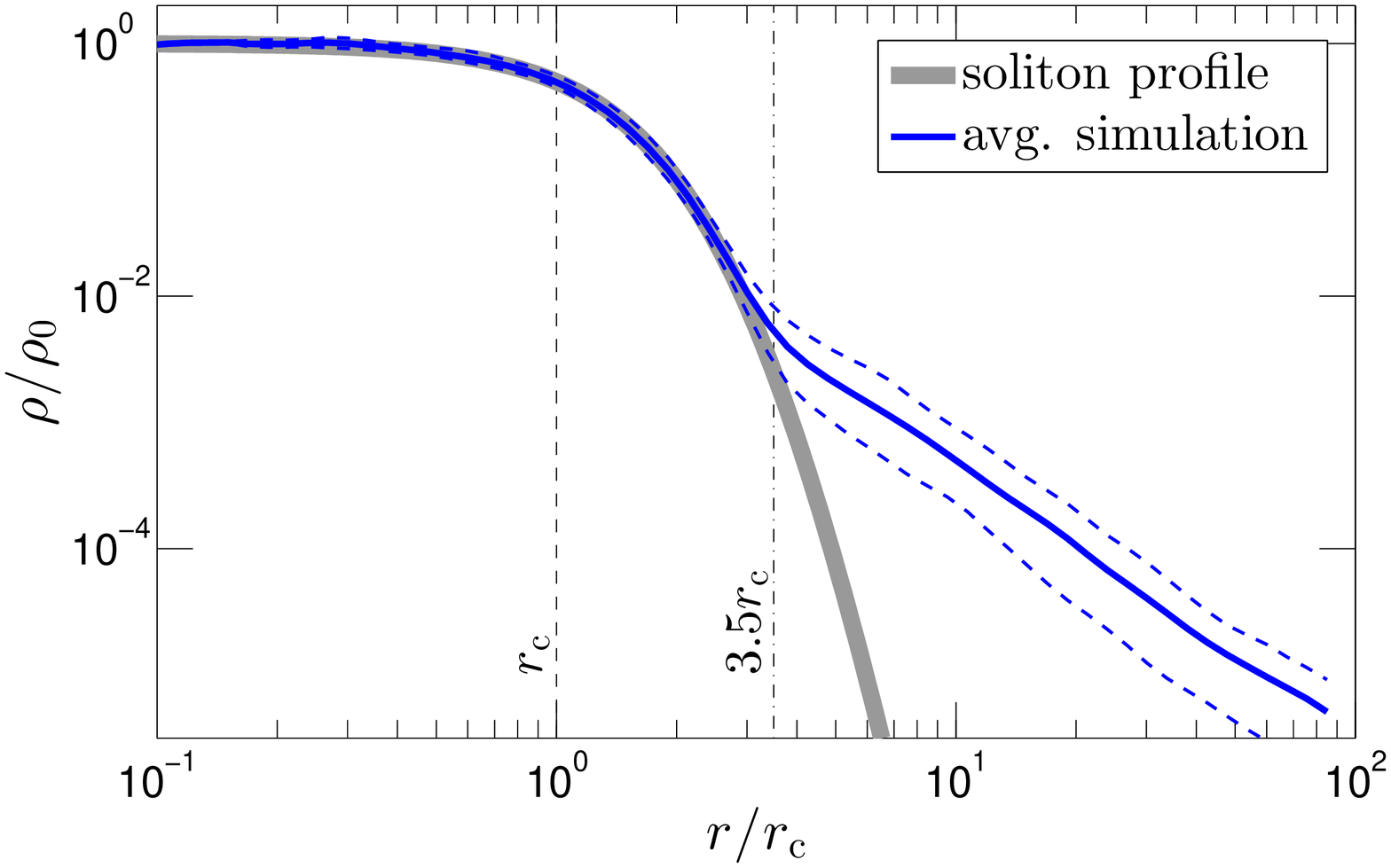} 
\end{center}
\caption{\textit{Top.} The radial density profiles of the virialized haloes formed in our set of simulations. The inner profile is well characterised by a soliton, and the outer profile follows an $r^{-3}$ NFW-like drop with some scatter (see text for details). We highlight the simulation data for $3$ random haloes, and show the soliton fits (solid lines) for all $100$ simulated haloes. \textit{Bottom.} Scaled density profiles (independent of scaling symmetry $\lambda$) showing that the soliton profile is self-similar. The $5$ per cent and $95$ per cent quantile contours are shown for our sample of 100 simulations. The profiles are universal, with a break at $r\simeq 3.5r_{\rm c}$, beyond which there is a variation in the normalisation of the outer power-law slope, depending on the total mass of the halo.} 
\label{fig:profiles}
\end{figure}

\begin{figure}
\begin{center}
\includegraphics[width=0.47\textwidth,angle=0,clip=true] {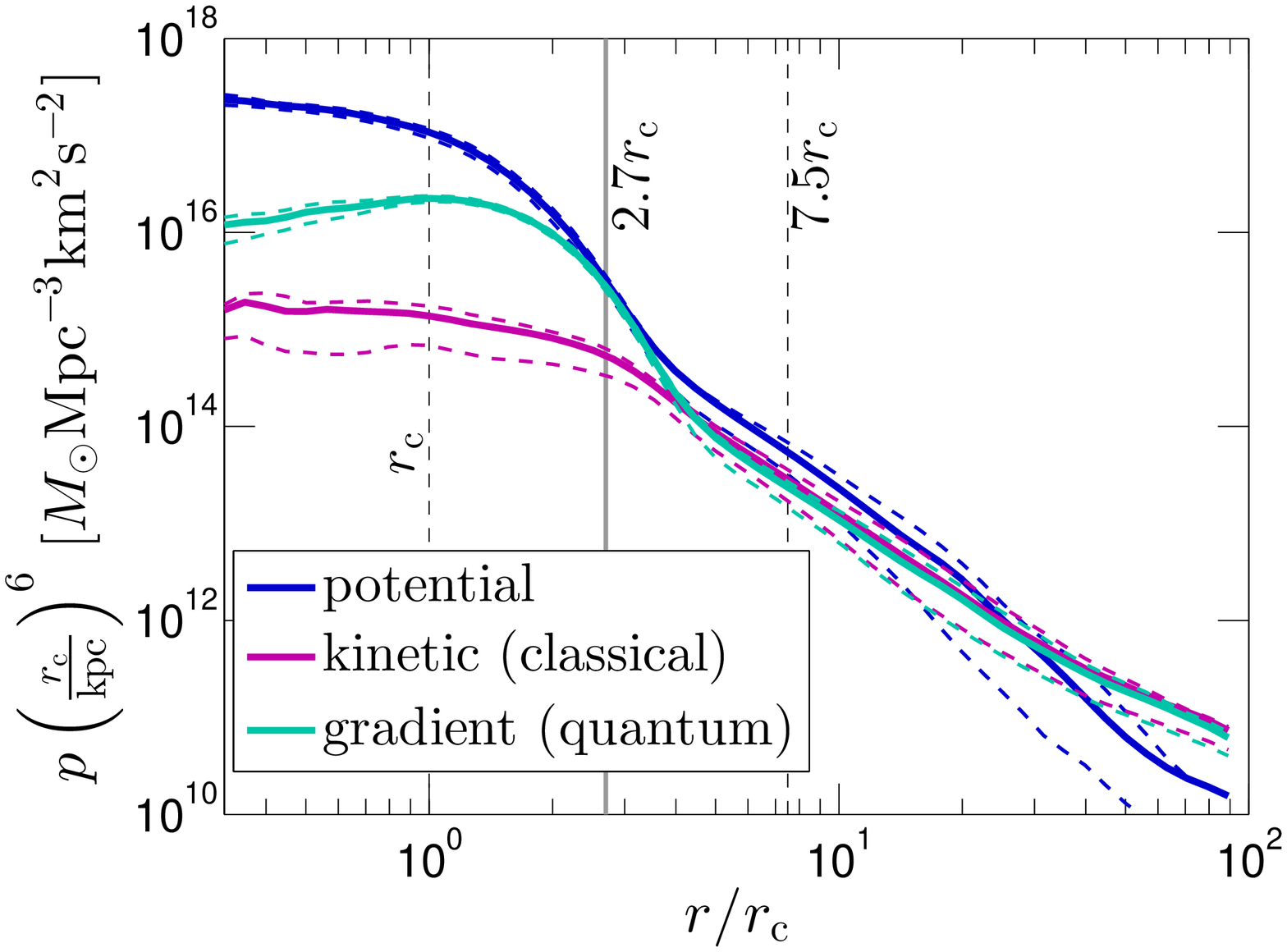} 
\includegraphics[width=0.47\textwidth,angle=0,clip=true] {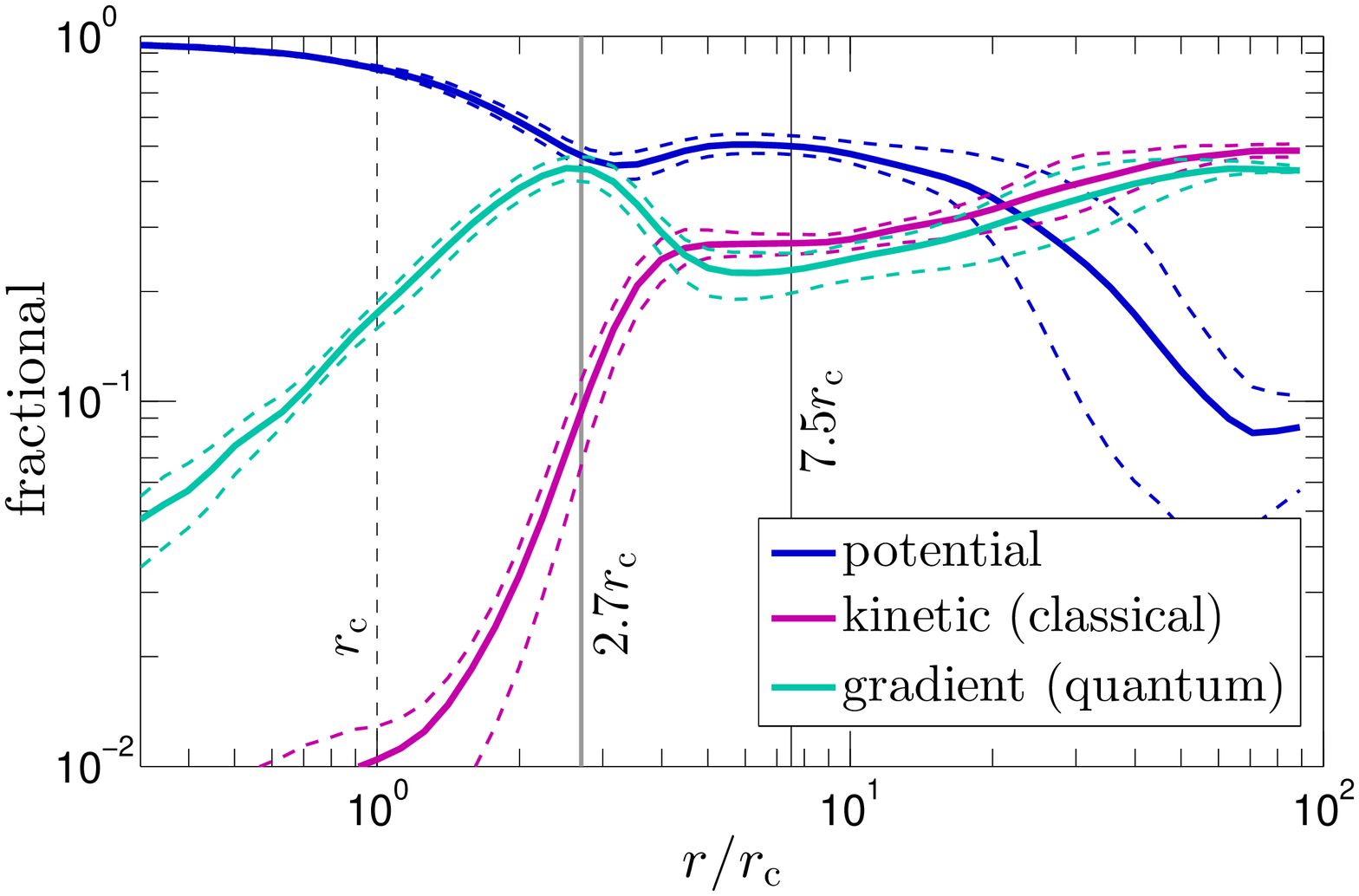} 
\end{center}
\caption{\textit{Top.} The radial energy density profiles (potential $\frac{1}{2}\rho \lvert V \rvert$, classical kinetic $\frac{1}{2}\rho v^2$, and quantum gradient $\frac{\hbar^2}{2m^2}(\nabla\sqrt{\rho})^2$; note the units are units of pressure, and the potential energy density is a positive quantity in this log-log plot) of virialized haloes in our simulations. The $25$ per cent and $75$ per cent quantile contours are shown for our sample of $100$ simulations. The vertical dashed line represents the radius parameter $r_{\rm c}$ of the central soliton core. The profiles are normalised by a factor of $r_{\rm c}$ due to the scaling symmetry of the SP equations. Inside the soliton core, the structure is supported by the quantum gradient energy density which peaks at $\sim 2.7 r_{\rm c}$. Classical kinetic energy is minimal in the center signaling a \textit{cold} core. In the $r^{-3}$ NFW-like outskirts a tight equipartition is established between the three types of energy densities. It is at very large radii the potential energy contribution becomes subdominant (due to boundary conditions).
\textit{Bottom.} Same as top figure, but instead showing the \textit{fractional} contribution of each of the energy density components as a function of radius; i.e., the core and global properties of BECDM haloes are linked. The vertical line at $\sim 7.5\,r_{\rm c}$ shows the location where the contribution of the total kinetic energy is exactly half of the potential energy; additionally, the fractional kinetic energies are comparable.}
 
\label{fig:energyProfile}
\end{figure}

\subsection{Simulation Setup}\label{sec:simsetup}

We simulate bound systems with various total energy $E$ and total mass $M$,
both of which are conserved in the total system which consists of a cubic box of $1~{\rm
Mpc}$ on a side. We enforce periodic boundary conditions so there is no loss of
energy or mass in the simulation. The periodic box allows us to account for
incoming waves from all directions, which is closer to the cosmological case
\citep{2014NatPh..10..496S} where waves from other haloes at larger distances
would also interfere with a given host halo. A similar setup was used in
\citet{2014PhRvL.113z1302S} where they studied idealised merger simulations
with a smaller suite. Notably, we find a different fundamental scaling between
the core mass and global quantities of the system, and will give some possible
explanations for this variation below. Additionally,  our study is different
and complementary to \citet{2016PhRvD..94d3513S}, where they analysed mergers
of binary solitons in a finite volume where no wave reflection at the
boundaries was allowed, they use the same boundary conditions as analytical
studies making their results more comparable to those expected for isolated
systems.

The primary variable that defines our systems is the invariant (under $\lambda$) $\lvert E \rvert/M^3$. For the initial condition at $t=0$, we randomly place between $4$ and $32$ cores with randomly selected soliton radii $r_{\rm c}\in[8,50]~{\rm kpc}$, allowing for multiple mergers at any time. We assume no phase offsets in the wavefunction between the cores.

The simulation uses internal units of $[{\rm L}] = {\rm Mpc}$, $[{\rm M}] =
M_\odot$, $[{\rm v}] = {\rm km}~{\rm s}^{-1}$. Our highest resolution
simulations have a resolution of $512^3$ cells and are run until a time
$t_{\rm end}=10$ (internal units; the physical units are scalable and we rescale all the simulations presented in the paper to a Hubble time), which gives more than sufficient time for the dark
matter structure to virialize over many dynamical timescales. The smallest
final cores found in our simulations are resolved by at least 4 cells per linear dimension.  In our
results and analysis, we rescale the simulations with the scaling parameter
$\lambda$ (Eqn.~\ref{eqn:scaling}), chosen so that the total simulation time is
the Hubble time $t_{\rm H}$, hence $\lambda=\sqrt{t_{\rm H}/t_{\rm end}}=26$. 
We stress that the scaling symmetry is a very important feature of the SP
equations and our results can be rescaled to other halo masses in a
straightforward way. By the scaling symmetry, lower mass haloes take a longer
time to virialize. In the internal units of our code, the typical halo masses
of our simulations fall into the range of few $\times 10^7~M_\odot$ to few
$\times10^9~M_\odot$, although we can scale our solutions by using $\lambda$.
However, for generality, most of our results are presented in a way in which
the scaling is factored out and are thus directly applicable to any mass halo.
The core properties in the subsequent sections are analysed at the final time
of one Hubble time.
The simulations were scaled to such a time to demonstrate that halos of such masses become virialized well before the Hubble time.

For the simulations, we used a boson mass of $m = 2.5\times 10^{-22}~{\rm eV}$,
the same as the fiducial value in \citet{2016PhRvD..94d3513S}.  We stress once again, that the
results may be scaled with mass $m\to\alpha m$ according to Eqns.~\ref{eqn:alphaScaling1} and \ref{eqn:alphaScaling}.

Fig.~\ref{fig:snaps} shows a volume rendering of the density field of one of
our simulations at $4$ different times ($t=0,0.1t_{\rm H},0.2t_{\rm H},t_{\rm
H}$), plotted with \textsc{yt} \citep{2011ApJS..192....9T}.  As the solitons
merge, they interfere quantum mechanically and create waves and interference
patterns in the fluid.  Shown also are the energy components of the fluid in
the box as a function of time. The total energy $E$ is conserved to within a
few percent by our method.  The quantum gradient energy dominates over the
classical kinetic energy, and provides the support against gravitational
collapse on small scales.  The system is in approximate virial equilibrium: $0
\simeq 2\langle K_\rho + K_v\rangle + \langle W\rangle$.

\begin{figure}
\begin{center}
\includegraphics[width=0.47\textwidth,angle=0,clip=true] {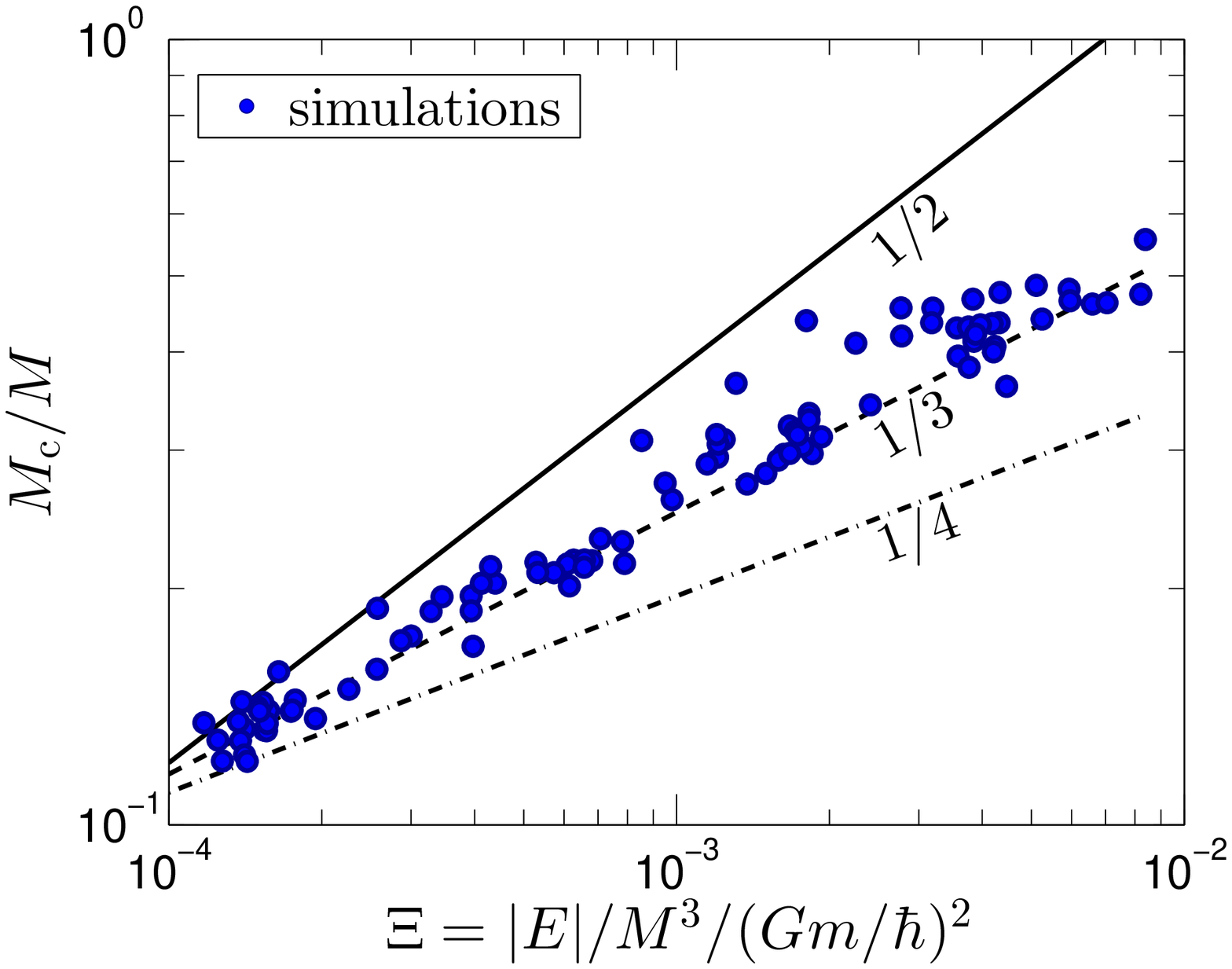} 
\includegraphics[width=0.47\textwidth,angle=0,clip=true] {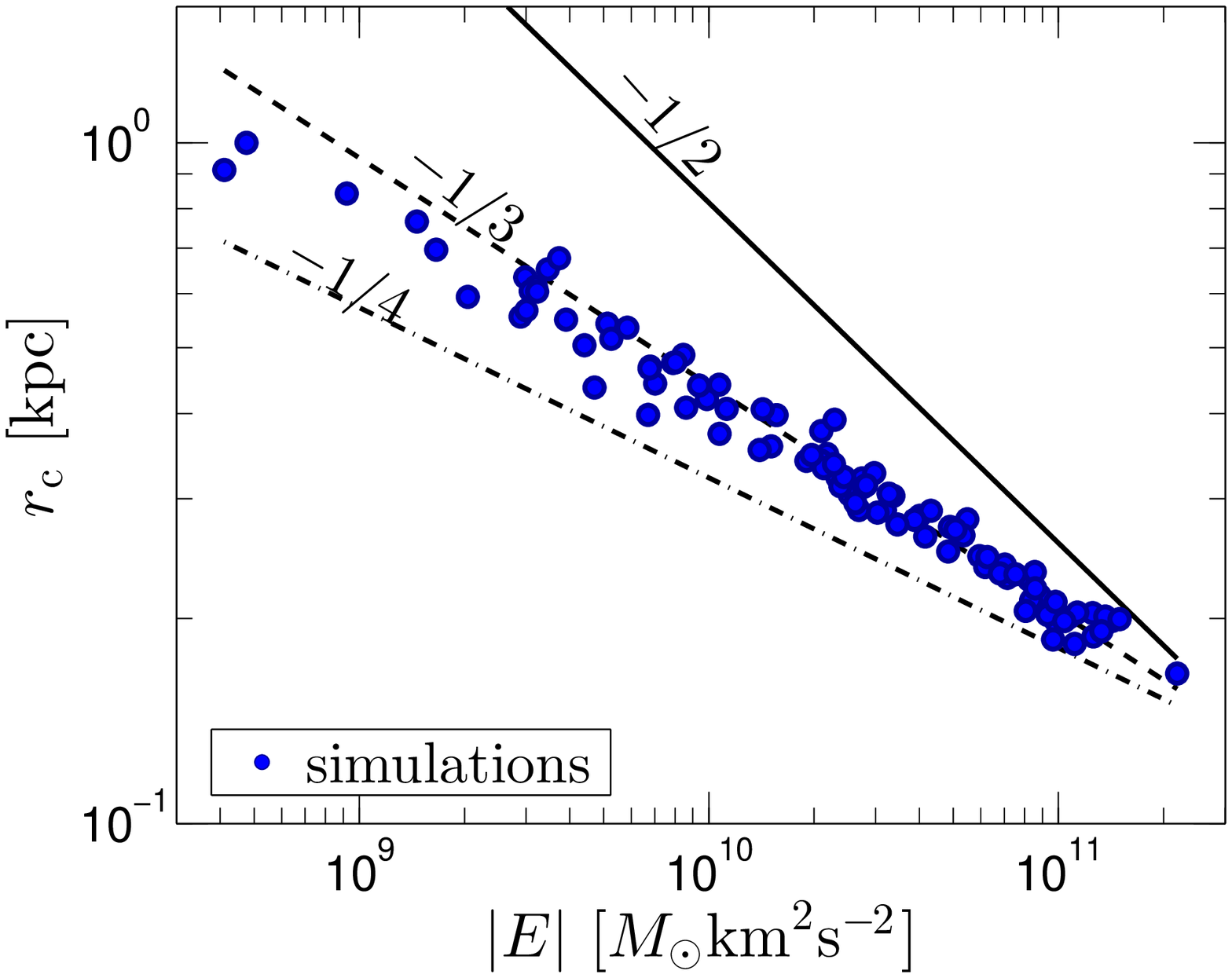} 
\end{center}
\caption{\textit{Top.} The (normalised) soliton core mass $M_{\rm c}/M$, where $M$ is the total (halo) mass, is tightly correlated with the invariant quantity $\Xi \equiv \lvert E \rvert/M^3/(Gm/\hbar)^2$ for all virialized haloes in our simulations, exhibiting a power-law slope of $1/3$. \textit{Bottom.} In dimensionful form, a consequence of the relation is that the core radius is correlated with the total energy of the system with a power-law slope of $-1/3$. This implies that the soliton core energy traces the halo total energy, where the halo energy may be estimated as: $\lvert E \rvert \sim GM^2/R_{\rm h}$.
}
\label{fig:mc}
\end{figure}

\section{Halo profiles}\label{sec:profiles}

The primordial dark matter solitons in our merger simulations all lead to the formation of final haloes with a central soliton core, which is well-described by the analytic form found in previous BECDM simulations in the literature (Eqn.~\ref{eqn:fit}). These inner solitons are consistent with stability studies where it was shown to be an attractor solution under perturbations. The soliton core has a flat central density followed by a sharp drop in density (as steep as $r^{-16}$). Importantly, there is only a single parameter that characterises the cores: the core radius, or, equivalently, the core mass as the two are related by
\begin{equation}
M_{\rm c}(r_{\rm c}) =  3.59\times 10^7 \left(\frac{2.5\times 10^{-22}~{\rm eV}}{m}\right)^2 \left(\frac{r_{\rm c}}{\rm kpc}\right)^{-1} ~M_\odot
\end{equation}
Once the core radius is chosen, the normalisation of the soliton core is determined by the balance between gravitational collapse and the quantum mechanical support: there is no freedom to choose the normalisation. More massive soliton cores are smaller in radius. 

Interestingly, the outer profile of the simulated virialized BECDM haloes is found to follow a $r^{-3}$ power-law,  
similar to the outer profile of an NFW halo in CDM; the break occurs universally at about the soliton size $r_{\rm soliton}\simeq 3.5r_{\rm c}$. 
Fig.~\ref{fig:profiles} shows the profiles of each of the $100$ haloes, along with the soliton core fits. After normalising the density profiles to their central density, we observe a unique soliton profile with some scatter beyond $r_{\rm soliton}$ attributed to the turbulent behavior. 
It is important to note that because we assume periodic boundary conditions and since mass is conserved, the system is continuously being perturbed by the reflecting waves that do not attenuate at the edges of the box, the latter precludes reaching the equilibrium configurations that are found analytically, where perturbations from the outer regions cease in a finite time and the systems may reach a dominant mode or configurations with multiple excited states \citep{Matos2007,rob13,med15,bernal17,ber10,ruff69}. 
Given our assumptions and in virtue of the stability arguments for BECDM haloes, it is then expected that the central ground state soliton will be the only mode that remains after a long evolution and the system has reached approximate virial equilibrium. 

The radial energy densities of virialized haloes (potential, classical kinetic, quantum gradient) are plotted in Fig.~\ref{fig:energyProfile} to highlight general properties. The soliton core is supported against gravitational collapse by the quantum pressure tensor, which is expected from the analytic description of steady-state soliton cores. The classical kinetic energy is sub-dominant in the core. The core is stable and protected against disruption from turbulent perturbations, which start appearing at $r \sim 3.5 r_{\rm c}$, just where the quantum pressure and kinetic energies become comparable. Equipartition in the three energy densities is seen, however, in the outer parts of the profile, in the $\rho\propto r^{-3}$ region (i.e., the three energy densities follow the same radial profile up to a constant factor); for larger radii the potential energy becomes subdominant (due to boundary effects). Equipartition can be a characteristic feature of turbulence, as is the case here for this dissipationless fluid. The breakdown of the energy densities is found to be universal across all our simulations.

We note that in the region of the stable soliton core, the wave function actually has a time dependent
phase \citep{2004PhRvD..69l4033G}:
\begin{equation}
\psi_{\rm soliton}(r,t) = e^{-i \hat{\gamma} \lambda_0^2 t/(\hbar m c^2)}\sqrt{\rho_{\rm soliton}(r)}
\end{equation}
where $\hat{\gamma} = -0.69223$, and
\begin{equation}
\lambda_0 = \left( \frac{\rho_0}{m^2c^4/(4\pi G \hbar^2)} \right)^{1/4}
\end{equation}
This corresponds to a characteristic period of:
\begin{eqnarray}
T &=& 2\pi \left(\lvert\hat{\gamma}\rvert \lambda_0^2/(\hbar m c^2)\right)^{-1} \\
&=& 6.9\times  10^8 \left(\frac{2.5\times  10^{-22}~{\rm eV}}{m}\right) \left(\frac{r_{\rm c}}{1~{\rm kpc}}\right)^2~{\rm yr}
\end{eqnarray}
In our simulations we do see oscillations of the core mass/radius of order a few percent with this characteristic frequency (also seen as oscillations in the energy components in Fig.~\ref{fig:snaps}).
This may result from the intrinsic phase of the stable soliton being slightly perturbed by interfering constructively and destructively with the turbulent/chaotic uncorrelated phases of the surrounding turbulent medium (see Section~\ref{sec:pspec}).
We note that the soliton core stays smooth and free of substructure at all times in this oscillation.

\begin{figure}
\begin{center}
\includegraphics[width=0.47\textwidth,angle=0,clip=true] {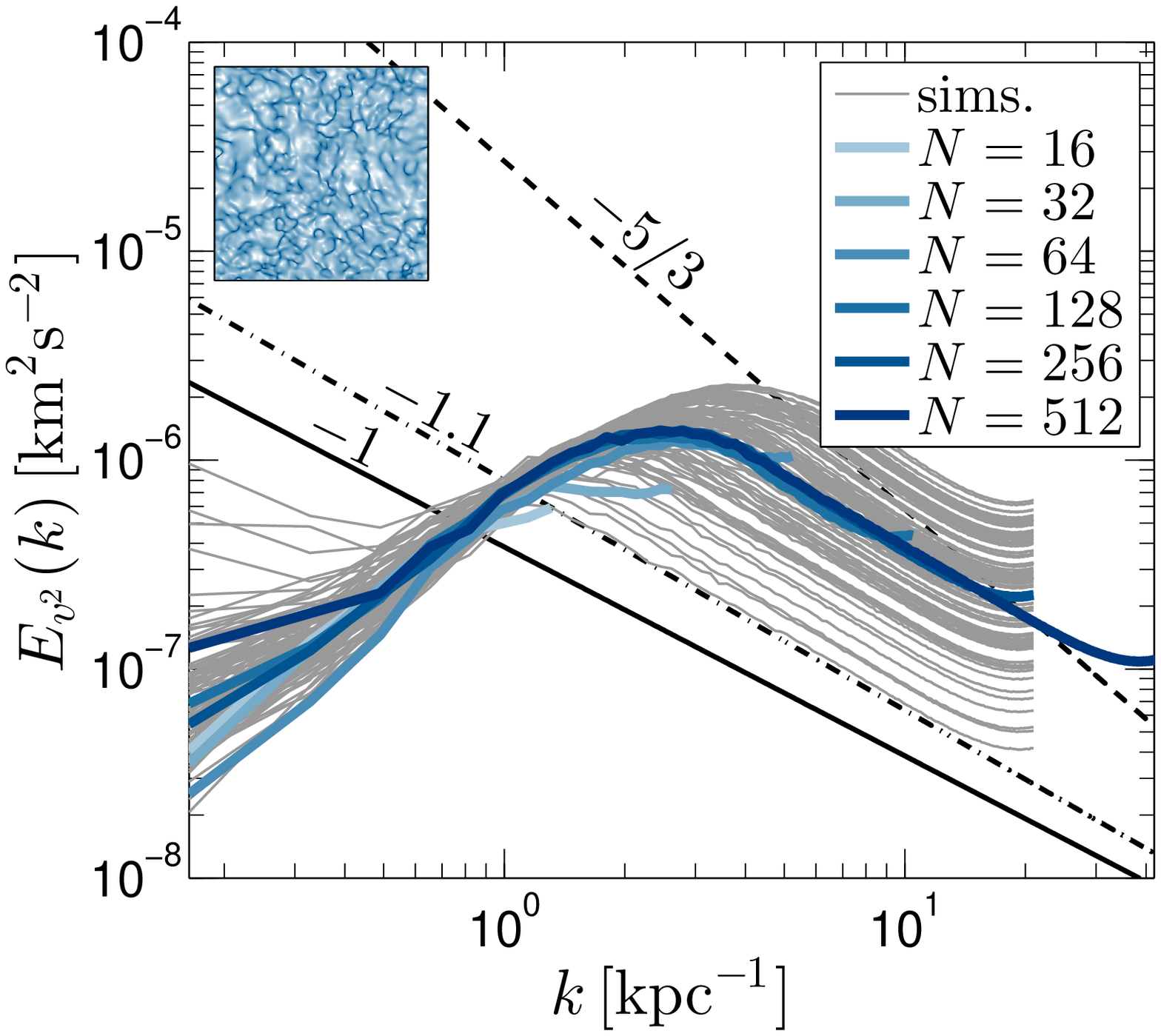} \\
\includegraphics[width=0.47\textwidth,angle=0,clip=true] {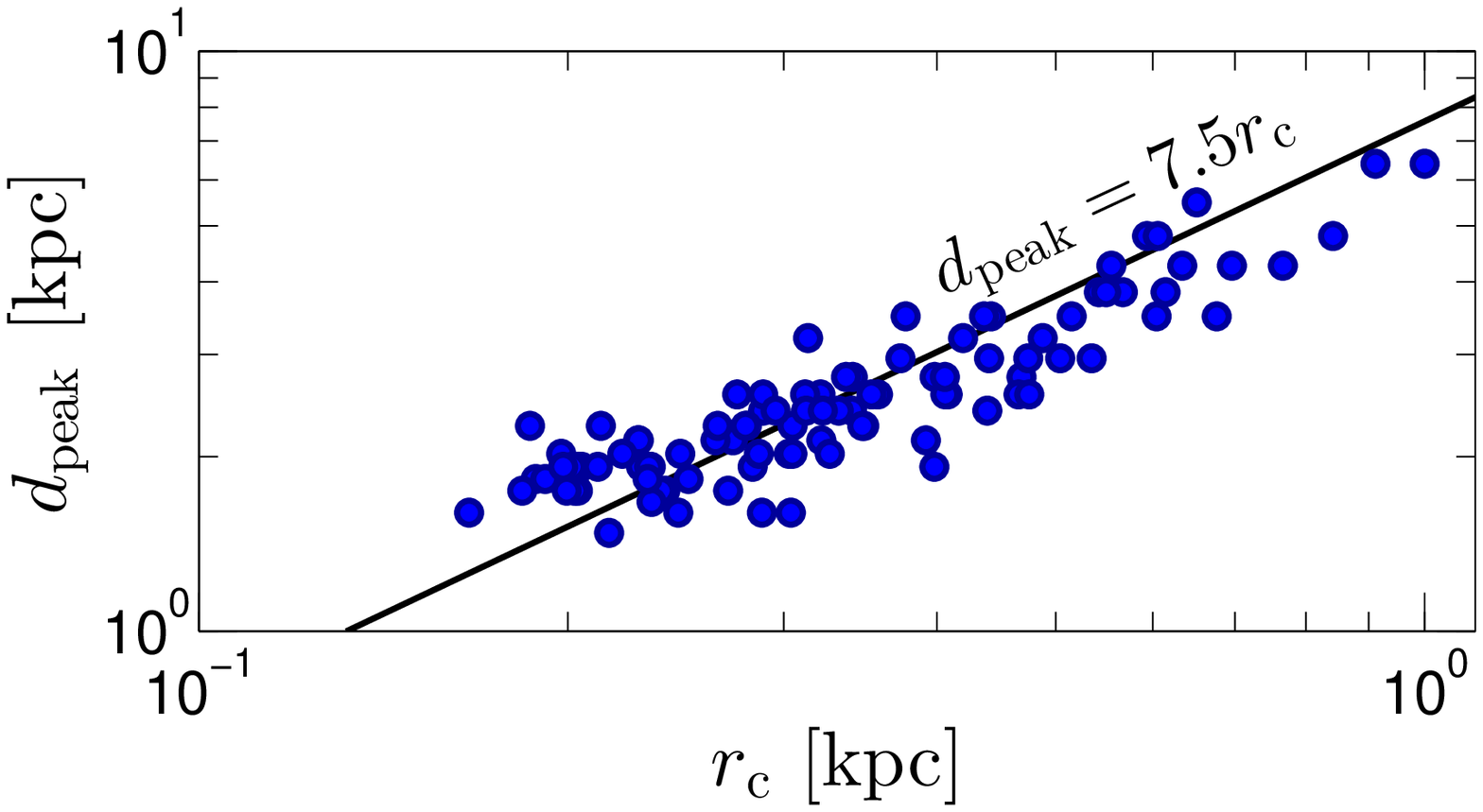} 
\end{center}
\caption{\textit{Top.} Velocity power spectra of the $100$ simulations. The power spectra follow a $k^{-1.1}$ relation, similar to a `thermally driven' counterflow $k^{-1}$ spectrum with a bump at intermediate (inter-vortex) scales seen in some other BEC systems, rather than a $k^{-5/3}$ Kolmogorov power-law that would arise from mechanical driving at the largest scales. The plot also shows the power-spectra calculated for various resolutions ($N^3$) for one of the simulations, indicating that the slope is well-converged. The inset shows a slice of the field $v$ (velocity norm) in the box, which is homogeneous throughout the domain.
\textit{Bottom.} Plot of the correlation between turbulent peak power scale $d_{\rm peak}$ and core size $r_{\rm c}$. Also shown (black line) is our fit $d_{\rm peak}$=7.5$r_{\rm c} \sim 2 r_{\rm soliton}$.
} 
\label{fig:pspec}
\end{figure}

\begin{figure*}
\begin{center}
\includegraphics[height=1.72in,angle=0,clip=true] {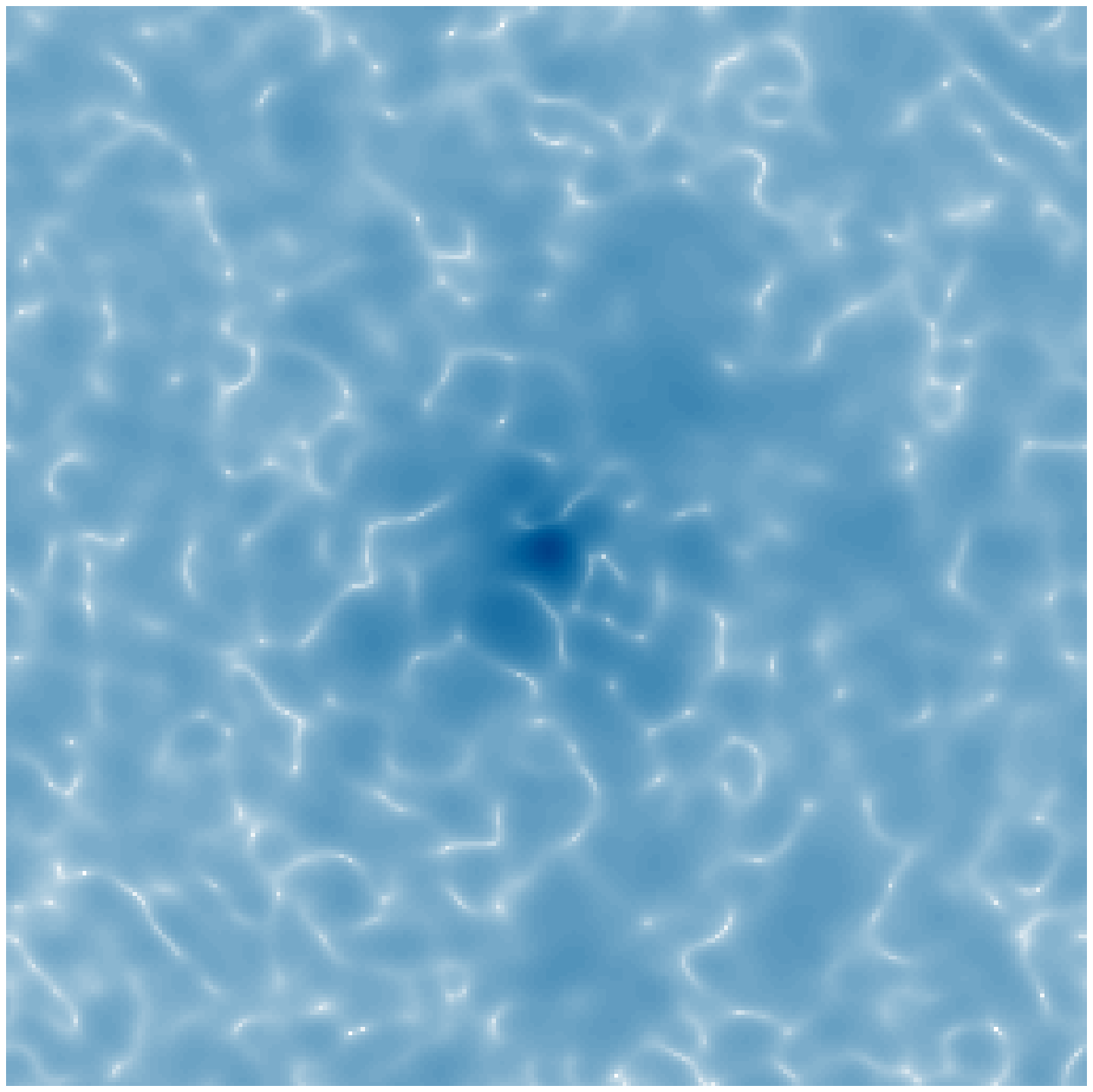} 
\includegraphics[height=1.72in,angle=0,clip=true] {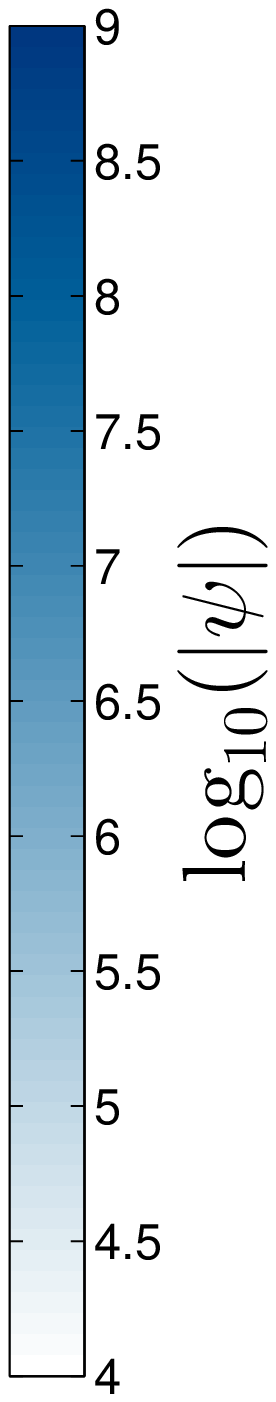} 
\includegraphics[height=1.72in,angle=0,clip=true] {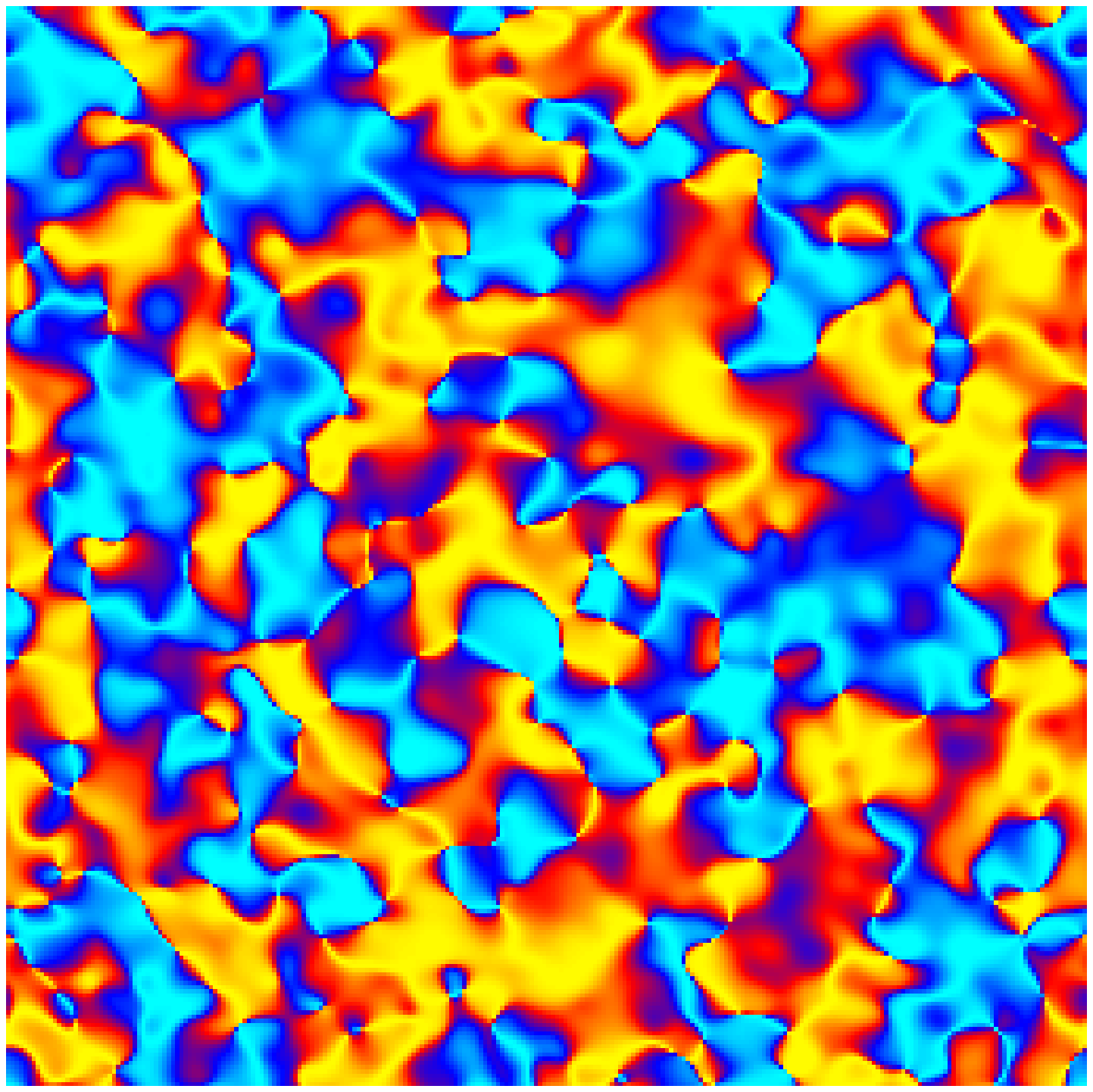} 
\includegraphics[height=1.72in,angle=0,clip=true] {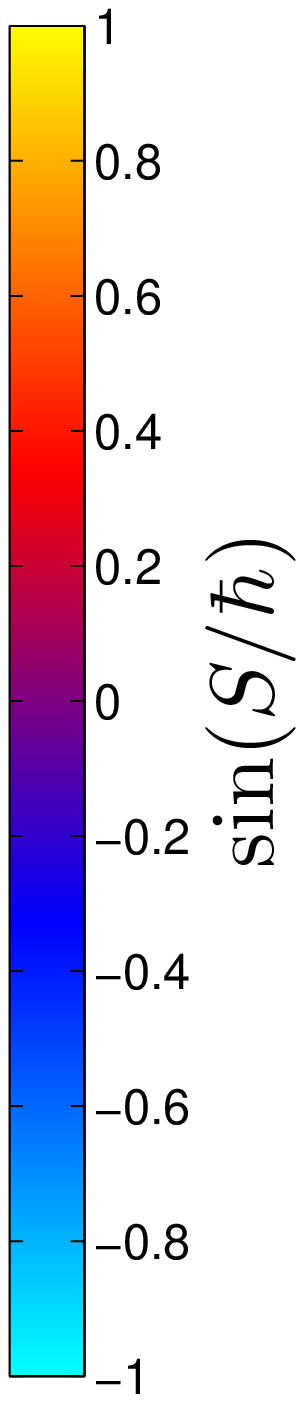} 
\includegraphics[height=1.72in,angle=0,clip=true] {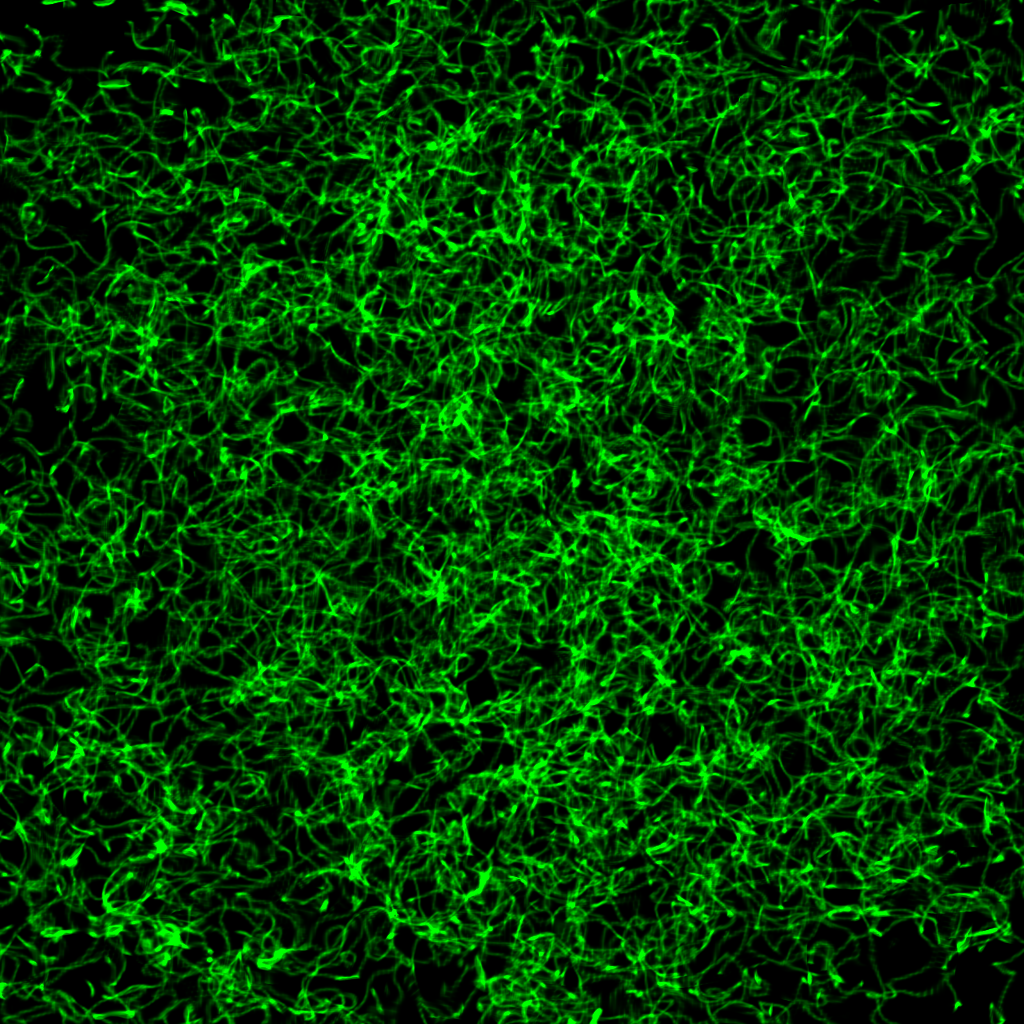} \\ \bigskip 
\begin{tabular}{cc}
\includegraphics[height=1.2in,angle=0,clip=true] {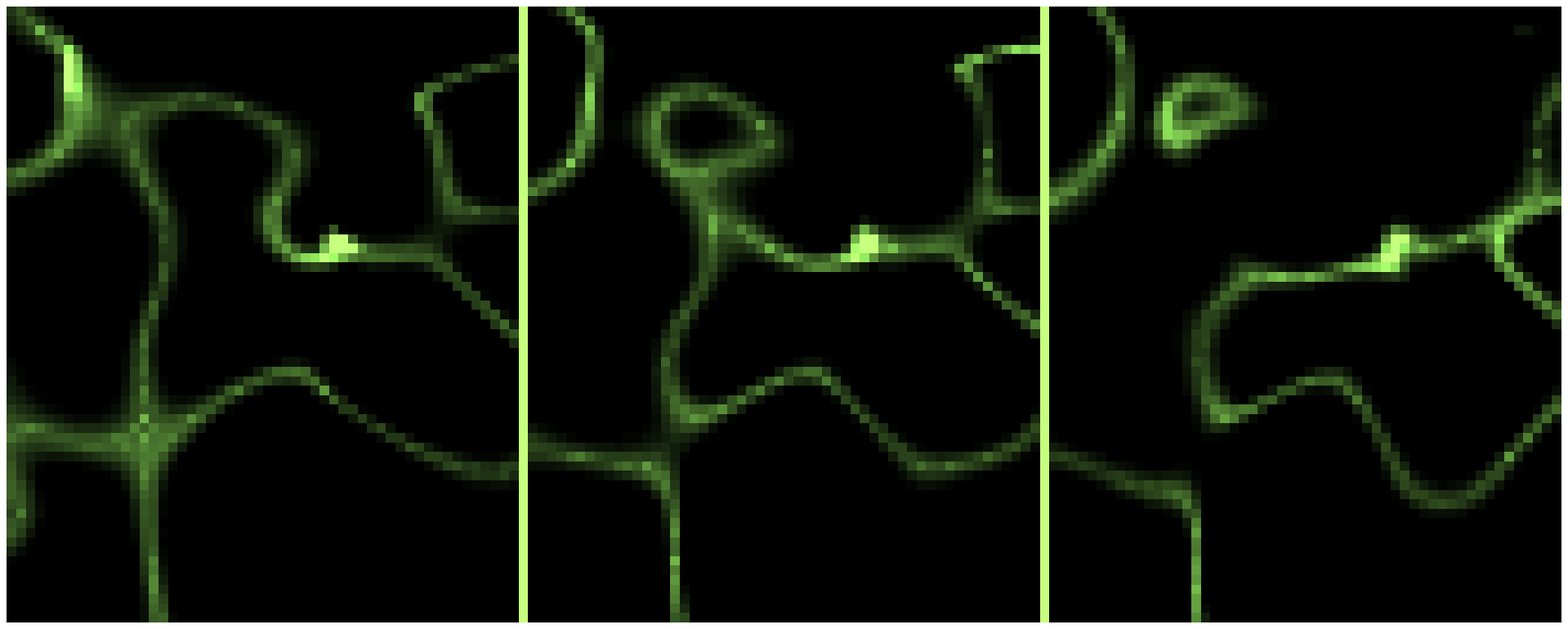} &
\includegraphics[height=1.2in,angle=0,clip=true] {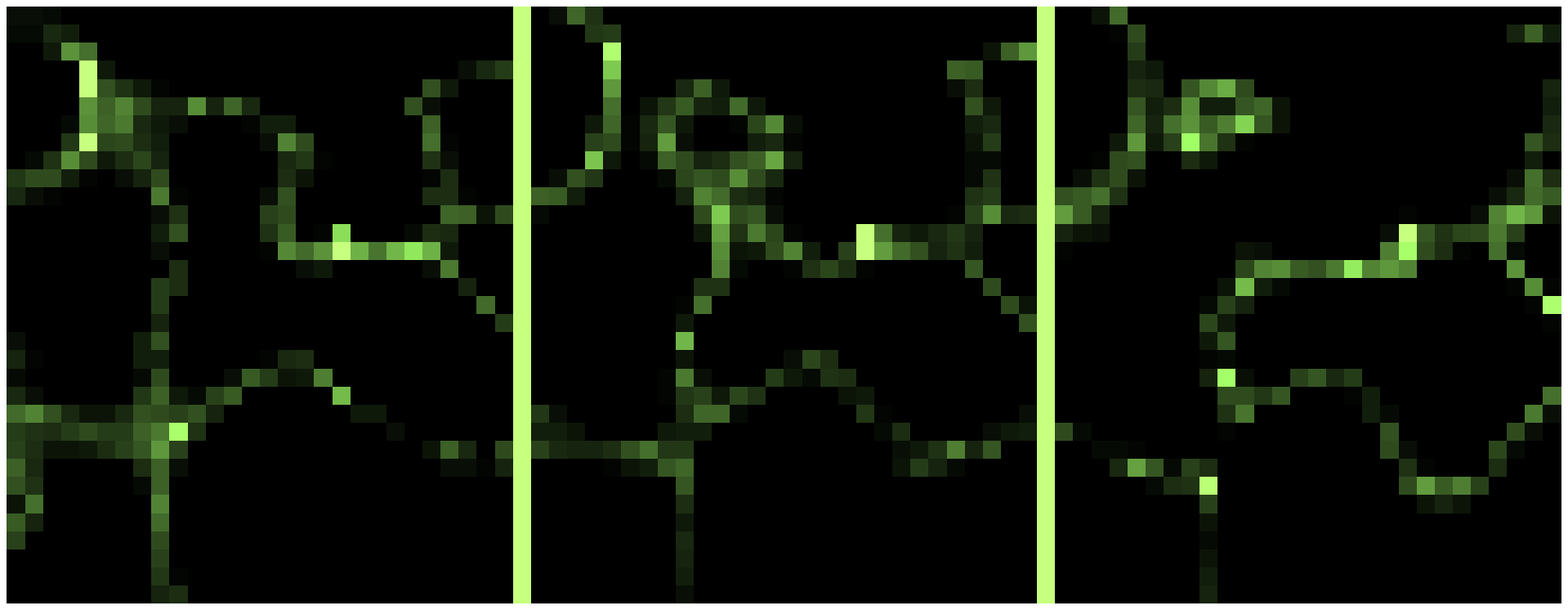} \\
resolution $512^3$ & resolution $256^3$
\end{tabular}
\end{center}
\caption{Vortex lines in the virialized BECDM haloes. Shown are the 2D slice of the absolute value of the wavefunction $\lvert\psi\rvert$ (\textit{top left panel}), the wavefunction phase $S$ (\textit{top mid panel}), and a 3D volume rendering of the vortex lines in the $1$ Mpc box domain (\textit{top right panel}). Vortex lines occur where there is a discontinuity in the phase, which occurs at $\lvert\psi\rvert\rightarrow 0$. The system of vortex lines contains all the vorticity in the superfluid, and their reconnection drives turbulence.
The \textit{bottom panel} shows a zoom-in on reconnection events in a $\sim 5~{\rm kpc}$ region in the box at $t=t_{\rm H}$ for simulation resolutions $512^3$ and $256^3$. The individual snapshots separated in time by $\Delta t=t_{\rm H} / 1000$. The individual reconnection events (hence reconnection rate) are converged, owing to the exponential spatial convergence properties of the spectral method.
} 
\label{fig:vcores}
\end{figure*}

\section{Soliton Core Mass Scaling}\label{sec:core}

We study how the resulting soliton core mass scales with the other fundamental parameters of the system, namely the total mass $M$ and energy $E$. More precisely, $M$ and $E$ here refer to the total mass and energy in the 1 Mpc box, as they are scale invariants in the SP system.
\cite{2014NatPh..10..496S,2014PhRvL.113z1302S} claim a relation
$M_{\rm c}\propto (\lvert E\rvert/M)^{1/2} $.
This is an interesting relation, because the left-hand side scales as the inverse of the soliton radius  $r_{\rm c}^{-1}$ and the right hand side is proportional to the halo velocity dispersion $\sigma_{\rm h}$, 
which gives the relation $r_{\rm c}\sigma_{\rm h} \sim 1$, 
a nontrivial type of non-local uncertainty principle.
However, \cite{2016PhRvD..94d3513S} point out that this relation may not be fundamental to the system, but may in fact be biased, due to the scaling symmetry of the fluid. Both sides of $M_{\rm c}\propto (\lvert E\rvert/M)^{1/2} $ scale as $\lambda$, the scaling parameter, so a linear relation would simply be found between these two parameters due to the scaling parameter alone, with no fundamental origin. \cite{2016PhRvD..94d3513S} recommended to look for fundamental relations by looking for relations between scale-free invariants, such as $\lvert E\rvert/M^3$. The authors do not find a single scaling that fitted their sample of mostly two-body collision simulations, possibly due to the small range of energies sampled by their suite and the strong dependence on the mass ratio of the two merging solitons. 
 
In our 100 simulations of virialized multi-body mergers, essentially characterised by a single parameter $\Xi \equiv \lvert E \rvert/M^3/(Gm/\hbar)^2$ set by the initial mass and energy (we have assumed no net angular momentum), we do find a fundamental relation between core mass $M_{\rm c}$ and $\Xi$.
\begin {equation}
M_{\rm c}/M \simeq 2.6 \Xi^{1/3}=2.6 \biggr( \frac{\lvert E \rvert}{M^3 (Gm/\hbar)^2}\biggl)^{1/3},
\label{eq:mcvsm}
\end{equation}
which  reproduces our simulations spanning two orders of magnitude in $E$, as shown in Fig.~\ref{fig:mc}.
More precisely, a numerical fit to the data yields $M_{\rm c}/M \simeq 2.89^{\pm 0.26} \Xi^{0.346^{\pm 0.013}}$, or $M_{\rm c}/M \simeq 2.63^{\pm 0.04} \Xi^{1/3}$ if the slope is fixed to $1/3$, where $1$-$\sigma$ errors are reported. \cite{2016PhRvD..94d3513S} do not find this result in their mainly two-body merger simulations, because the two-body results are sensitive to the mass ratio and the total angular momentum. More importantly, \cite{2016PhRvD..94d3513S} used sponge boundary conditions, in which a quasi steady-state solution is not reached; rather, mass and kinetic energy escape at the boundaries, which intrinsically change the total mass $M$ in Eqn.~(\ref{eq:mcvsm}). These boundary conditions are closer to those used in analytical solutions of isolated haloes, while our simulations have closer resemblance to a cosmological scenario. Different boundary conditions and varying angular momentum in the simulations could potentially be a source of the differences in the results, but such detailed comparison is outside the scope of this work.

The relation $M_{\rm c}/M\propto \Xi^{1/3}$ simplifies to
$M_{\rm c} \propto \lvert E \rvert^{1/3}$ implying that the soliton core traces the total energy of the system. Equivalently, the relation may be written also as:
$r_{\rm c} \propto \lvert E \rvert^{-1/3}$
or
$\lvert E_{\rm core} \rvert \propto \lvert E \rvert$.
Essentially, the relationship tells us there is tight coupling between core and global properties as seen in Fig.~\ref{fig:mc}. These relations suggest that global quantities of virialized haloes such as the total mass or energy are enough to estimate the expected core size, this will be important in cosmological simulations of BECDM where structures appear at several halo masses. 

We could also get an estimated relation for the fundamental parameters in the following way: suppose from potential theory, that the gravitational potential at the
center of the soliton is proportional to the gravitational potential 
at the center of the halo. We derive what this means for the scaling relation of the form $M_{\rm c}/M\propto \Xi^{\eta}$.
For the soliton core, $M_{\rm c} r_{\rm c} \sim \hbar^2/(Gm^2)$, so the velocity dispersion is:
$v_{\rm c}^2 \sim GM_{\rm c}/r_{\rm c} \sim G^2 M_{\rm c}^2 m^2/\hbar^2$.
For the halo, 
$v_{\rm h}^2 \sim GM/R_{\rm h}$ and $\lvert E \rvert \sim GM^2/R_{\rm h}$.
Assuming a constant ratio $v_{\rm c}/v_{\rm h} \equiv \mu$, one may deduce:
\begin{equation}
 M_{\rm c} /  M \sim \mu \Xi^{1/2}
\end{equation}
The power $\eta = 1/2$ signifies a constant core to halo velocity dispersion ratio.
In our simulations, however, we observe $\eta = 1/3$,
meaning that there is weak mass, energy dependence in the velocity dispersion ratio, namely
$\mu \propto \Xi^{-1/6}$.

We note that a power of $\eta = 1/2$
is derived analytically from self-similarity of the potential of the core and halo, which basically
is the assumption: $M_{\rm c}/r_{\rm c} \sim M/R_{\rm h}$.
On the other hand, $\eta = 1/3$
can be derived analytically from self-similarity of the energy of the core and halo: $M_{\rm c}^2/r_{\rm c} \sim M^2/R_{\rm h}$, which may be a better assumption for the strongly coupled turbulent system found in our simulation. This assumption leads to the observed weak mass, energy dependence of $\mu$.

It has also been suggested, in the literature, that $M_{\rm c}/M$ essentially follows from the mass 
loss in subsequent binary mergers \citep{2017PhRvD..95d3519D}. That work suggests $M_{\rm c}\propto M^{1.44(\beta-1)}$ where $\beta\leq 1$ is the descendant-to-originator mass-fraction.
Assuming the halo radius scales approximately as $R_{\rm h} \propto M^{1/3}$, 
then Eqn.~\ref{eq:mcvsm} implies $M_{\rm c}\propto M^{5/9}$, corresponding to $\beta=0.69$.
This is consistent with the result $\beta\sim 0.7$ of \cite{2016PhRvD..94d3513S,2017PhRvD..95d3519D}.

\section{Velocity Power Spectra}\label{sec:pspec}

We compute the 1D radial superfluid energy spectrum $E_{v^2}(k)$ defined by \cite{2012PhRvL.109t5304B}:
\begin{equation}
E_{v^2}(k) = \frac{1}{L_{\rm box}^3} \int\frac{1}{2}\lvert \mathbf{v}\rvert^2 \,d\mathbf{x}= \int E_{v^2}(k)\,dk.
\end{equation}
BEC systems without self-gravity but with the non-linear self-interaction term (Gross-Pitaevskii equations)
are known to reproduce a classical Kolmogorov scaling $E_{v^2}(k)\propto k^{-5/3}$ if mechanically driven on the box scale, and a shallower $E_{v^2}(k)\propto k^{-1}$ turbulent cascade in the counterflow regime for large $k$ beyond the inter-vortex length-scale `bump' and no power on the largest scales \citep{baggaley2012thermally}.

Fig.~\ref{fig:pspec} shows the energy power spectrum $E_{v^2}(k)$ for our simulations. The simulations themselves show sustained chaotic motions (stable kinetic energy with time, equipartition) and a homogeneous filamentary distribution of $v$ (which traces out the vortex lines in the fluid; Fig.~\ref{fig:vcores}). No turbulence appears inside the soliton. There is no power on the largest spatial scales, as expected, due to lack of large-scale driving, limited by the simulation box size. The simulations all show a well converged power-law relation of $E_{v^2}(k)\propto k^{-1.1}$, closer to the thermally-driven counterflow analog Gross-Pitaevskii system from condensed matter physics than the mechanically driven one, and we also observed a maximum mode that carries most of the energy in the turbulent medium. This energy power spectrum is characteristic of isotropic turbulence where small modes dominate the turbulence. 
As seen in Fig.~\ref{fig:pspec}, we find that with our highest resolution we can capture the scale where the spectrum peaks, very low resolution could lead to missing the peak mode $k_{\rm peak}$ and result in a lack of homogeneous turbulence in the simulation. 

In the bottom panel of Fig.~\ref{fig:pspec}, we show the relation between the scale $d_{\rm peak} = 2\pi/k_{\rm peak}$ where $E_{v^2} (k)$ peaks for each halo and the core size of the corresponding soliton. We find $d_{\rm peak} \sim 7.5r_{\rm c}$, this corresponds to the region where the potential energy density contributes equally to the kinetic energy density (Fig.~\ref{fig:energyProfile}). Notice that this wavelength $d_{\rm peak} \sim 2 r_{\rm soliton}$, which is the total width of the soliton and it is the typical scale where most of the interference is happening, since $r_{\rm soliton}$ is much smaller than the box, this explains the appearance of the homogeneous turbulence throughout the box.

Evidence for the existence of turbulence everywhere in the domain comes from the identification of vortex lines in our simulations (Fig.~\ref{fig:vcores}). These filamentary structures are a source of turbulence in a quantum fluid. Vortex lines are degenerate locations in the fluid that have a discontinuity in the differentiability of $\psi$ and have $\lvert\psi\rvert \to 0$. They contain all the vorticity in the fluid as the velocity field must be curl-free elsewhere. Turbulence persist at all times in the box since the system is closed and the total angular momentum ($\textbf{L}=\textbf{0}$) is conserved. The existence of vortex lines is a necessary condition for quantum turbulence, and their reconnection creates Kelvin waves that drive the turbulent motions.
Fig.~\ref{fig:vcores} shows a slice of the magnitude and phase of $\lvert\psi\rvert$. We see clear evidence of $\lvert\psi\rvert \to 0$ filamentary structures which correspond with discontinuities in the phase of $\psi$. 
The figure also shows a zoom-in on the network of reconnection events homogeneous throughout the domain (except inside the soliton code).
The reconnection events (hence reconnection rate) are numerically converged, as the figure shows the same events are identified for simulation resolutions $512^3$ and $256^3$ at the Hubble time.
The excellent convergence is due to the exponential spatial convergence properties of the spectral method.

The turbulent structure is seen everywhere in the domain, except, of course, inside the soliton core, which is protected from velocity fluctuations because it is a stable soliton solution.
Both the soliton width and the turbulent structures (peak of the velocity power spectrum) are $\approx 2 r_{\rm soliton}$, determined by the de Broglie length-scale of the system, the only length-scale in the problem.

\begin{figure}
\begin{center}
\includegraphics[width=0.47\textwidth,angle=0,clip=true] {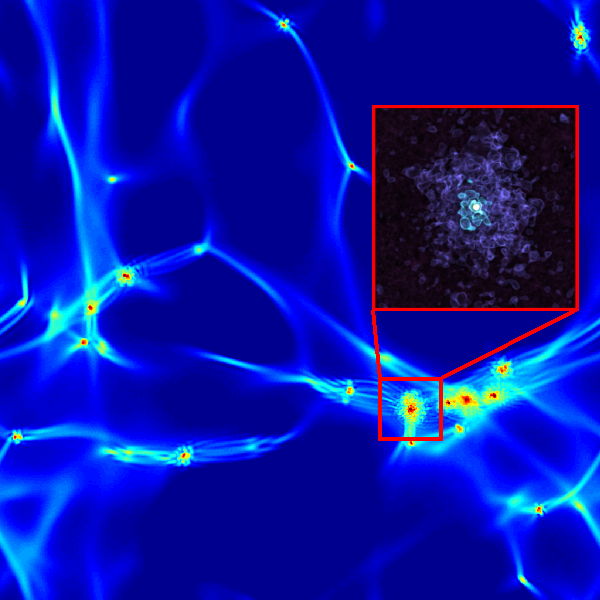}
\end{center}
\caption{A cosmological simulation with BECDM at $z=4$ ran with the {\sc arepo} code; an example of upcoming simulations for paper II (Mocz et al. in prep.). The inset shows the result of a simulation from this work. The figure adds cosmological context to our simulations. Our simulations are representative of the result of virialized cosmological mergers. Turbulence in the halos can be seen in the cosmological simulation as well.
} 
\label{fig:context}
\end{figure}

\section{Concluding Remarks}\label{sec:conc}

We have carried out $100$ simulations of virialized BECDM halo cores with periodic boundary conditions to
characterise their properties, which are largely universal and independent of
initial condition details. Merging multiple haloes with initially cuspy or cored profiles
both lead to the formation of stable soliton cores at the centers of BECDM
haloes. Our simulation setup provides a useful numerical laboratory for statistically studying the final product of the relaxation process of BECDM haloes with a wide range of total energies. The
structure of the resulting dark matter haloes depends primarily on the total mass
and energy of the system. The haloes form a stable soliton core with a turbulent
$r^{-3}$ NFW-like outer profile and (in the absence of angular momentum) are
characterised by a single dimensionless invariant:
$\Xi \equiv \lvert E \rvert/M^3/(Gm/\hbar)^2$. Contrary to previous works, we find that for all of our haloes the core mass of the 
soliton scales with this quantity $\Xi$ as $M_{\rm c}/M\propto \Xi^{1/3}$, which implies
$M_{\rm c} \propto \lvert E \rvert^{1/3} \Rightarrow \vert E_{\rm core} \rvert
\propto \lvert E \rvert$. Properties of the soliton at the centers of haloes
are therefore tightly linked to the global halo properties. 

Soliton core profiles are described by a single parameter (the core mass, or
equivalently the core radius, as the two are related by
$M_{\rm c}\propto r_{\rm c}^{-1}$). The size/mass of the core that forms traces
the total energy of the entire virialized dark matter halo. We found that the typical soliton size is $~3.5r_{\rm c}$,  beyond this radius in the BECDM profile the haloes are found to be turbulent, exhibiting a filamentary distribution of vortex lines that form during merging events and are sustained and reconnect to drive
turbulence. No turbulence is seen inside the soliton. Equipartition between the potential energy density, classical kinetic energy density, and quantum gradient energy density is seen in the outer
core, maintaining a continuously perturbed medium. The turbulence is characterised by a $k^{-1.1}$ power-law in the velocity power spectrum, characteristic of an isotropic turbulence in the fluid. We find this is because the dominant mode in the 1D superfluid velocity spectrum peaks at a scale twice the soliton radius, which is several times smaller than the total length of the system. We find that the suppression of turbulence inside the soliton and the existence of a maximum mode in the velocity power spectra with a scale equal to the soliton width, could explain why the typical scale for the granules in the density field of BECDM simulations is preferentially the soliton size. 

The cuspy halo profile universally found in $\Lambda$CDM
simulations \citep{1996ApJ...462..563N} has seen tension with observations that 
suggest core-like potentials at halo centers.
A recent example of such an observation is of SPT-CLJ2011-5228 \citep{collett2017}, a $z=2.39$ system
gravitationally lensed by a $z=1.06$ cluster along the line-of-sight. The inner
density profile falls with radius to the
power $-0.38\pm0.04$ ($1\sigma$ errors) out to a radius of
$270^{+48}_{-76}$~kpc. This shallow inner profile is in strong tension with our
understanding of relaxed cold dark matter haloes, where NFW predicts a $r^{-1}$
profile, and perhaps the flat slope is suggestive of a central soliton. 

An interesting application of the systems studied here may be in 
neutron star glitch statistics, as neutron star interiors may be modelled as a
superfluid by the Gross-Pitaevskii equations
\citep{2011MNRAS.415.1611W,2013MNRAS.428.1911W}. Glitches may 
originate from the turbulent nature of the fluid, along with the possible
intermittent nature of turbulence.

BECDM has been largely studied in the past and a number of independent constraints exist on the boson mass that would make up this type of dark matter, as outlined in the introduction. However, the analytical arguments ultimately need to be validated by numerical simulations to ensure that the analytic assumptions made in the derivations are valid, as was the case historically for the standard CDM scenario. 
Some of the analyses have suggested moderate tension in the
boson mass; however, only through full BECDM cosmological simulations (ultimately involving
full baryonic physics) we will be able to confirm these
assumptions and place tighter constraints on the boson mass. 
For example, Lyman-$\alpha$ constraints have not included quantum density fluctuations, which are present in the BECDM model and seen in our simulations

\subsection{Context and outlook}

The next step in our work will be to simulate the BECDM model coupled to
baryons in a fully cosmological setting (Mocz et al. in prep.) to address the
impact of baryons on the predictions of BECDM only simulations. We have
implemented the numerical method presented in this paper into the
{\sc arepo} code, so we will be able to run the axion dark matter
simulations fully-coupled to baryonic components with feedback in upcoming
work.

Fig.~\ref{fig:context} shows an example of a BECDM $1~{\rm Mpc}$ cosmological simulation ran
with the {\sc arepo} code, at redshift $z=4$, with resolution of $1024^{3}$ cells for a boson mass of $m = 2.5\times 10^{-22}~{\rm eV}$. The resolution of the simulation is $\sim 1$~kpc, enough to capture turbulence and the soliton cores.
We have highlighted in the
figure a virialized halo as a result of cosmological mergers. Turbulence is
also seen in the cosmological box at the intersections of cosmic web filaments where halos have merged, and our current simulations are able to
provide a high resolution characterisation of the phenomenon (i.e., the turbulent cascade is resolved over a broader range).  The BECDM
cosmological simulations, to be described in detail in Paper II, are created with a
realistic axion power spectrum at $z\sim 100$ and we aim at using them to explore the effect of varying the boson mass, and to study the coupling with baryonic physics.

\section*{Acknowledgments}
This material is based upon work supported by the National Science Foundation
(NSF) Graduate Research Fellowship under grant no. DGE-1144152 (PM). PM is
supported in part by the NASA Earth and Space Science Fellowship (NNX12AB23C). VHR is supported by a UC MEXUS-CONACYT fellowship. PM would like
to thank  Jerry Ostriker, Alma Gonzalez, Luis Urena-Lopez, and Mikhail Medvedev for valuable
discussions, and Sauro Succi for reading of an earlier
version of this manuscript. The computations in this paper were run on the
Odyssey cluster supported by the FAS Division of Science, Research Computing
Group at Harvard University. MV acknowledges support through an MIT RSC award, the support of the Alfred P. Sloan Foundation, and support by NASA ATP grant NNX17AG29G. MBK acknowledges support from the NSF (grant
AST-1517226) and from NASA through ATP grant NNX17AG29G and HST theory
grants (programs AR-12836, AR-13888, AR-13896, AR-14282, and AR-14554) awarded
by the Space Telescope Science Institute (STScI), which is operated by the
Association of Universities for Research in Astronomy (AURA), Inc., under NASA
contract NAS5-26555. JZ acknowledges support by a Grant of Excellence from the Icelandic Research Fund (grant number 173929-051).

\bibliography{mybib}{}

\bsp
\label{lastpage}
\end{document}